\documentclass{aastex63}

\received{November 2, 2020}
\accepted{AJ, Jan 2021}
\submitjournal{AAS Journals}

\shorttitle{Binary Exoplanet Host Stars}
\shortauthors{Howell et al.}
\graphicspath{{./}{figures/}}

\begin{document}

\title{Speckle Observations of TESS Exoplanet Host Stars: \\
Understanding the Binary Exoplanet Host Star Orbital Period Distribution}

\correspondingauthor{Steve B. Howell}
\email{steve.b.howell@nasa.gov}


\author[0000-0002-2532-2853]{Steve~B.~Howell}
\affiliation{NASA Ames Research Center, 
Moffett Field, CA 94035 USA}

\author[0000-0001-7233-7508]{Rachel A.~Matson}
\affiliation{U.S. Naval Observatory, 3450 Massachusetts Avenue NW, Washington, D.C. 20392, USA}

\author[0000-0002-5741-3047]{David R.~Ciardi}
\affiliation{NASA Exoplanet Science Institute Caltech/IPAC Pasadena, CA 91125 USA}

\author[0000-0002-0885-7215]{Mark E. Everett}
\affiliation{NOIRLab, 950 N. Cherry Ave., Tucson, AZ 85719 USA}

\author[0000-0002-4881-3620]{John~H.~Livingston}
\affiliation{Department of Astronomy, University of Tokyo, 7-3-1 Hongo, Bunkyo-ku, Tokyo 113-0033, Japan}

\author[0000-0003-1038-9702]{Nicholas J. Scott}
\affiliation{NASA Ames Research Center, Moffett Field, CA 94035, USA}

\author[0000-0003-2159-1463] {Elliott P. Horch}
\affiliation{Department of Physics, Southern Connecticut State University, 501 Crescent Street, New Haven, CT 06515, USA}

\author[0000-0002-4265-047X]{Joshua N.\ Winn}
\affiliation{Department of Astrophysical Sciences, Princeton University, 4 Ivy Lane, Princeton, NJ 08544, USA}




\begin{abstract}
We present high-resolution speckle interferometric imaging observations
of TESS exoplanet host stars using the NN-EXPLORE NESSI instrument the at the 3.5-m WIYN telescope. Eight TOIs, that were originally discovered by Kepler, were previously observed using the Differential Speckle Survey Instrument (DSSI).
Speckle observations of 186 TESS stars were carried out and 45 (24\%) likely bound companions were detected. This is approximately the number of companions we would expect to observe given the established 46\% binarity rate in exoplanet host stars.
For the detected binaries, the distribution of stellar mass ratio is consistent with that of the standard Raghavan distribution and may show a decrease in high-$q$ systems as the binary separation increases. The distribution of binary orbital periods, however, is not consistent with the standard Ragahavan model and our observations support the premise that exoplanet-hosting stars with binary companions have, in general, wider orbital separations than field binaries. We find that exoplanet-hosting binary star systems show a distribution peaking near 100 au, higher than the 40-50 au peak that is observed for field binaries. This fact led to earlier suggestions that planet formation is suppressed in close binaries.
\end{abstract}



\section{Introduction} \label{sec:intro}

Our team has been carrying out high resolution speckle imaging of stars for which transit-like signals have been detected by the planet-finding missions Kepler \citep{2011ApJ...728..117B}, K2 \citep{2014PASP..126..398H}, and TESS \citep{2015JATIS...1a4003R}. High resolution imaging has proven useful for determining whether the signals are produced by planets or one of the various ``astrophysical false positives'' that plague wide-field transit surveys \citep{2011AJ....142...19H}. For those stars that do turn out to have transiting planets, high-resolution imaging also helps to characterize the basic system properties. Our decade-long program has provided high spatial resolution observations of thousands of exoplanet host stars. The final reduced data products are deposited in the public NASA Exoplanet Archive ExoFOP\footnote{https://exoplanetarchive.ipac.caltech.edu}.

The ongoing TESS mission, and its predecessors Kepler and K2, identify planet candidates by simultaneously staring at many stars in the sky, collecting highly precise photometric time series for each star. For TESS, the light curves have either 2-min or 30-min sampling, depending on whether the star was prioritized by the TESS Science Team or the Guest Investigator program. The light curves are searched for transit-like dips in brightness, tell-tale signatures of exoplanets orbiting across the face of their alien sun. 

An ideal photometer would be able to isolate the light from each and every target star, in which case the observed fractional loss of light would be $(R_p/R_\star)^2$, the area of the planet's silhouette divided by the area of the stellar disk. However, because of the limited angular resolution of the telescopes, this simple interpretation is often not appropriate. Each TESS camera pixel subtends about 20 arcseconds, and the digital apertures that are defined to produce the photometric time series consist of many pixels. Multiple stars may be present in the aperture, one or more of which may be, for example, variable or an eclipsing binary. The signal of a deep eclipse, when combined with the constant light from the target star, may mimic an exoplanet-like signal. This and other stellar configurations can be troublesome \citep{2011AJ....142..112B}, requiring follow-up observations to confirm or validate transiting planets.

Given that about half of the stars harboring exoplanets are in binary or multiple star systems, knowledge of bound companions is critical in allowing a complete and proper characterization of the exoplanet properties as well as providing robust tests of planet formation and evolution scenarios. \citet{2019AJ....157..211M} used such information to make discovery predictions for high resolution imaging detections of bound companions for the TESS mission. ``Third light" contamination within the aperture reduces the transit depth, causing the analysis to yield an exoplanet of a smaller radius than it really is \citep{2015ApJ...805...16C}. Other effects as well come into play which can produce incorrect characterization of both the host star's properties \citep{2020ApJ...898...47F} and, with the incorrect planet radius, a skewed planet radius distribution and occurrence rates \citep{2018AJ....156..292T,2018AJ....155..244B} as well as improper mean density and atmospheric values \citep{2020FrASS...7...10H}.

Several studies of Kepler and K2 exoplanet host stars have found companion fractions of 40-50\% (e.g.~\citealt{2014ApJ...795...60H}; \citealt{2016MNRAS.455.4212D}; \citealt{2018AJ....156...31M}; \citealt{2018AJ....156..259Z}), consistent with solar-type stars in the solar neighborhood \citep{2010ApJS..190....1R}. However, other studies find fewer close binary companions around Kepler exoplanet host stars \citep{2016AJ....152....8K} and TESS planet candidate host stars \citep{2020AJ....159...19Z}. These studies find a deficit of close binary systems with projected separations less than $\sim$40 au.

\citet{2018AJ....156...31M} identified exoplanet candidate host stars from K2 that have stellar companions {\it{within}} 40 au based on the projected separation of the detected companion and the estimated distance to the system. To date, it has remained unclear if close binaries are able to host exoplanets and whether the formation and survival of a planetary system is possible under such conditions. For instance, planet formation in close binaries may depend not only on the presence of a stellar companion, but also on orbital parameters such as eccentricity and mutual inclination between the planetary system and the binary \citep{2016ApJ...817...80D}.
Discovering exoplanets that form and evolve in diverse physical characteristics which provide different dynamic interactions compared to our own Solar System pose many questions for the leading planet formation theories, especially for exoplanets residing in binary star systems \citep{2015pes..book..309T}.


TESS was launched in April 2018. After a few months of on-orbit check-out, it began to observe the southern sky in July 2018. Northern sky observations began in July 2019, and thus ground-based observations of northern sky TESS targets only began in earnest in late fall 2019. We present herein, the results of our first year of TESS high resolution speckle imaging follow-up using the NN-EXPLORE NESSI instrument at the WIYN 3.5-m telescope.

\section{Observations}
\label{sec:obs}

\subsection{Target Selection}

Starting in June 2019, we began follow-up observations of stars believed to host exoplanets discovered by the NASA TESS satellite during its second year of operation, a time period in which it surveyed the northern sky. A few preliminary equatorial targets were observed in June 2019 with the majority of the northern sky TESS targets being observed in October and November 2019. 

Using the mission's list of TESS Objects of Interest (TOIs) that are made public on ExoFOP\footnote{https://exofop.ipac.calteach.edu/tess/}, stars with robust software pipeline vetted transit-like signals, and additional community discovered exoplanet candidate host stars (known by their TESS Input Catalogue (TIC) designation) we observed 186 targets with NESSI at WIYN during the summer and fall of 2019. Our observation time was obtained through the NN-EXPLORE program\footnote{https://exoplanets.nasa.gov/exep/NNExplore/}, we ran a queue at WIYN, reduced the speckle interferometric data, and placed all our reduced data products, without a proprietary period, into the NASA ExoFOP. 

\subsection{WIYN Observations and Data Reduction}

Speckle observations presented in this paper were accomplished using the NESSI high resolution speckle imaging instrument \citep{2018PASP..130e4502S} mounted on the 3.5-m WIYN telescope located at Kitt Peak National Observatory. NESSI is a dual-channel imager using high-speed readout EMCCD detectors with plate scales of 0.0182 arcsec/pixel and a dichroic to split the optical light at $\sim$700 nm. Speckle images are obtained in a shutterless stream of 1000 images per set, each image being 40 ms in duration.
Depending on the target brightness, 3 or more sets are obtained in a row, each producing a simultaneous pair of blue and red images. The NESSI observations used a 562/40 nm blue filter and a 832/40 nm red filter.
The speckle images are stored as multi-extension FITS files (data cubes) of 1000, 40 ms images each. 

Resolved star systems produce a characteristic interferometric fringe pattern from which the separation, position angle, and delta magnitude can be determined through a modeling procedure. The raw FITS files 
are passed through our standard Fourier analysis pipeline \citep{2009AJ....137.5057H} in which the average power spectrum for each image is computed and summed. We next deconvolve the speckle transfer function through division by the power spectrum of a point source standard star (a nearby star that is observed at a similar time as the target star) and compute a weighted least-squares fit of a fringe pattern to the result. During this step, pixels in the Fourier plane that have low signal-to-noise and low-frequency values judged to be in the seeing disk are set to zero. In order to determine the highest probability quadrant location of the companion star, we compute a reconstructed image via bispectral analysis \citep{1983ApOpt..22.4028L}. Details of our data reduction techniques and error assessments are given in \citet{2011AJ....141...45H} and \citet{2011AJ....142...19H}.

\begin{figure}[t]
\centering
  \includegraphics[scale=0.55,keepaspectratio=true]{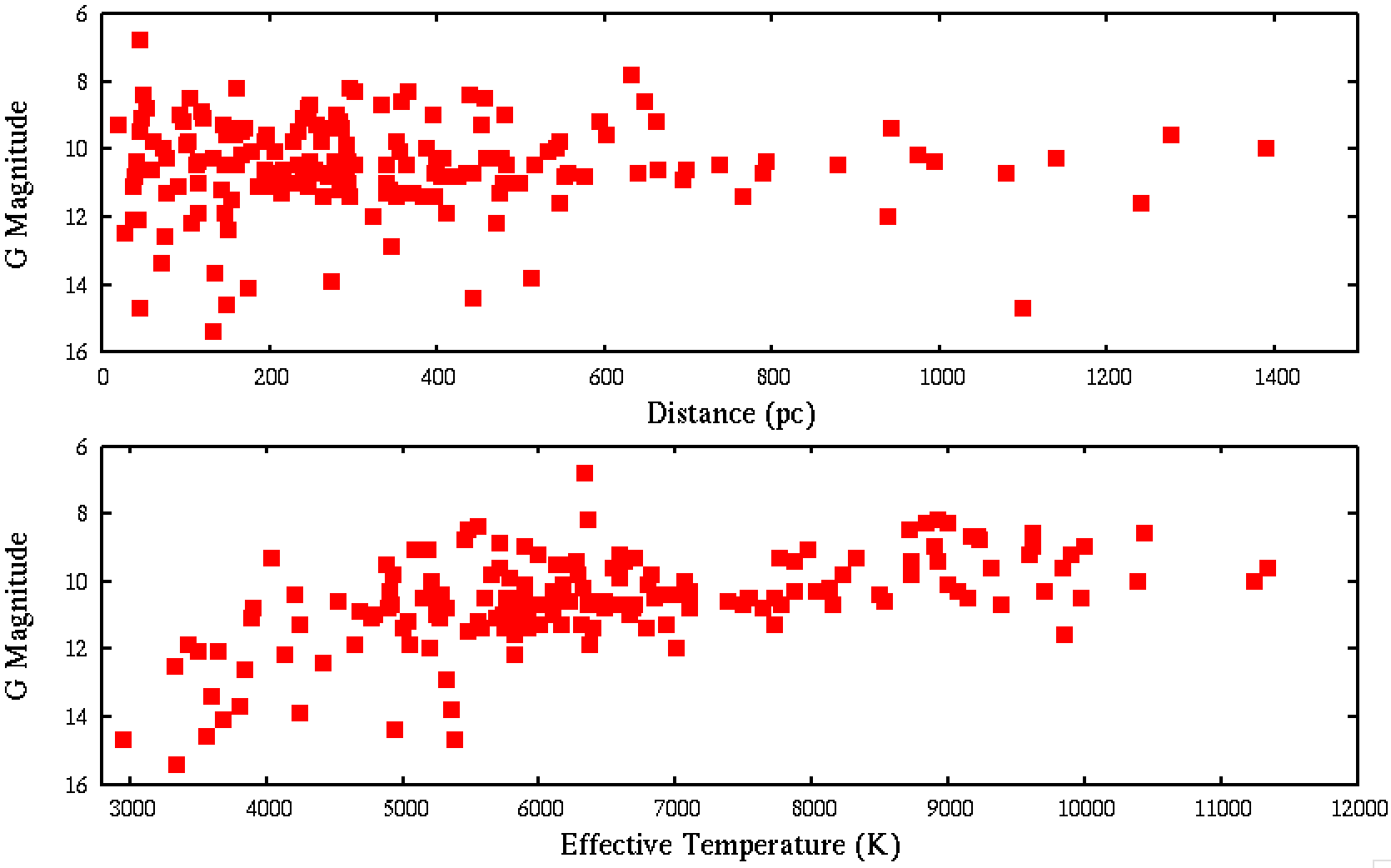}
  \caption{Properties of the TESS stars in our sample using ExoFOP TOI database values. Referenced to the DR2 Gaia apparent magnitude, these plots show the distribution of the distance and effective temperature within the sample. Most stars are near 10-11th magnitude, closer than 500 pc, and cooler than 7000K. A few more distant and hotter stars are not shown (See Table 1).}
  \centering
\end{figure}

Table 1 lists the TESS targets we observed and which will be discussed in this paper. In order to characterize each star we list in Table 1 some relevant stellar parameters; the Gaia magnitude, effective temperature of the star, and the Gaia determined distance as obtained from the ExoFOP archive TESS TOI table in October 2020. In all cases, we used the well-vetted ``default" or ``preferred" stellar parameters as given in the ExoFOP archive. Column 5 gives the date of observation, with the remaining four columns being the 5$\sigma$ $\Delta$mag contrast limit obtained in the observation at 0.2 and 1.0 arcsec in each band-pass.

Gaia parallaxes can be unreliable for close binaries as described in \citet{2018A&A...616A..17A} with a good discussion related to Gaia
parallexs and high-resolution imaging presented in \citet{2020AJ....159...19Z}. Close binaries resolved by Gaia, usually having separations of 0.7 arcsec or larger, are somewhat large by our standards. For unresolved binaries, \citet{2018A&A...616A..17A} states “…the astrometric quality of unresolved binaries with a small magnitude difference is not significantly different from that of single stars.” The uncertainties in the Gaia parallax values for resolved close binaries ($\ge$0.7 arcsec) manifest themselves in cases such as faint stars (fainter than G=$\sim$17), crowded sources (leading to duplicate entries), diffuse objects, and nearly resolved binaries spatially close to bright stars. We have six stars in Table 1 (TOIs 523, {\it{1152, 1162, 1163}}, 1172, and 1393) which fall into the duplicate and crowded/nearby bright star categories (See note to Table 1). Only three of these, those in italics in the above list, revealed a nearby star in our observations which Gaia also resolved. In all three cases, the companions are  widely separated from the primary star at 1.5 arcsec. While the distances to these three stars may have a larger than average parallax uncertainty, none of these wide companions are considered in the detailed analysis of this paper nor do they have any effect on its conclusions.

Speckle imaging provides angular resolutions to the diffraction limit of the telescope. For NESSI at WIYN, the inner working angle yields angular resolutions of 39 and 64 mas, providing spatial resolutions of 4-20 au (at 100-500 pc, 562 nm) and 6.2-31 au (at 100-500 pc, 832 nm). Eight of the 186 TESS TOIs presented herein were observed years ago at WIYN, as they were first detected as exoplanet host stars by the Kepler mission. 
These stars (identifiable in Table 1 by their date of observation) were observed using the Differential Speckle Survey Instrument (DSSI, \citealt{2009AJ....137.5057H}), which was the speckle imager used at WIYN from 2008 until NESSI was commissioned in 2016. The use of DSSI for exoplanet host star follow-up observations is described in \citet{2011AJ....142...19H}.
DSSI used similar filters to NESSI but with slightly different central wavelengths, 692 nm and 880 nm. All observing and reduction procedures were similar to those described above for NESSI.

Figure 1 shows two relevant properties of the TESS targets we have observed in this work, the distance and effective temperature of the sample as a function of the Gaia magnitude. Note that the stars cluster near 10-11th magnitude, 80\% are closer than 500 pc, and 75\% are cooler than 7000K. The closest stars (d$<$100 pc) represent the faintest and coolest stars in the sample.

Figure 2 shows the contrast range of the observations obtained for our targets as a function of their Gaia magnitude. For each band-pass, we note that the total range in contrast narrows, that is the delta magnitude limits become shallower, as the target star becomes fainter. This is a function of S/N in the Fourier summed images and why our standard observing practice is to use 3 or more images sets per observation depending on the target star magnitude\footnote{https://www.noao.edu/noao/staff/everett/nessi/Speckle\_with\_NESSI.html}.  Seeing too can have a similar effect in lowering the contrast of the final image. While the resolution of speckle imaging does not decrease with bad seeing, spreading the light out over a larger area both decreases the correlation of individual speckles across the image (having a correlation size of about 1 arcsec; \citealt{2019AJ....158..113H})
as well as making individual speckles harder to detect above any background sky noise during the 40 ms observation time. Both of these effects will reduce the interferometric signal in the data due to a lowering of the S/N in the final image\footnote{In normal CCD imaging, one could simply increase the exposure time to regain the S/N. However, longer exposure times yield ``seeing limited" imaging for which all high-resolution temporal and spatial interferometric information is lost.}.  
For the observations presented here, the majority were observed in seeing of 1.1 arcsec or better.

Examining Figure 2, it can be seen that the 562 nm bandpass keeps approximately the same sensitivity difference ($\sim$1 mag) between 0.2 and 1.0 arcsec over the entire magnitude range providing a contrast limit of 4-5 magnitudes at the bright end and 3-4 magnitudes near G=13. The 832 nm band-pass has deeper contrast limits at both reference angular separations. These observations reach a contrast of 7 magnitudes at the bright end (with a range of 2 magnitudes) and reach a contrast limit near 6 magnitudes with a range of 1 magnitude, near G=13.  Even though our EMCCDs have 10\% less detector QE at 832 nm vs.~562 nm, the atmospheric effects and speckle correlations for each target star are better at redder wavelengths. 
Greater contrast limits (up to 8 magnitudes) have been achieved at WIYN by obtaining more sets of speckle images because, to no one's surprise, the more time spent on a target the better S/N of the observation.

\begin{figure}[h!]
\centering
  \includegraphics[scale=0.55,keepaspectratio=true]{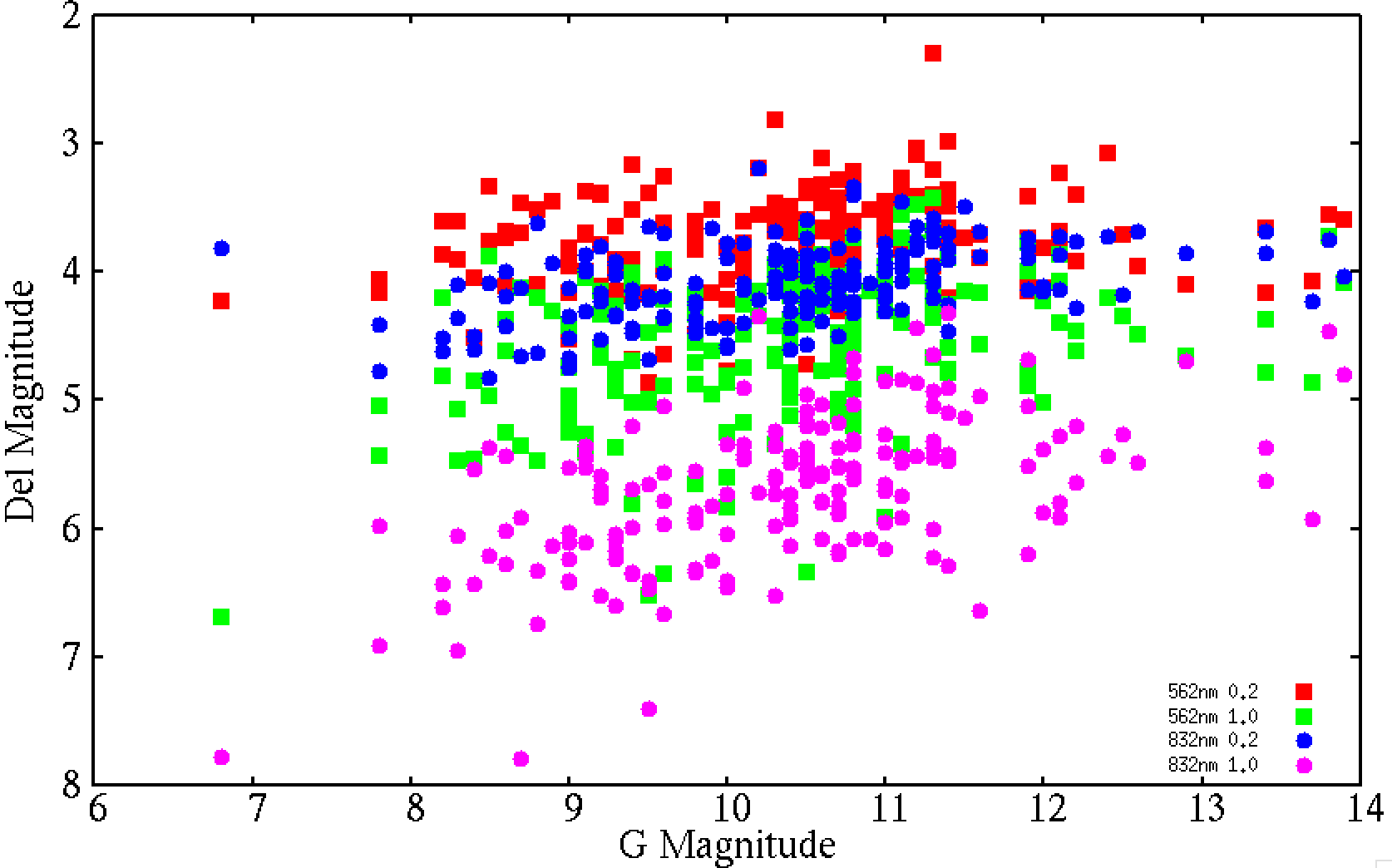}
  \caption{Speckle imaging contrast limits as a function of target star Gaia magnitude. The $\Delta$ magnitude contrast obtained at reference angular separations of 0.2" and 1.0" are shown as a function Gaia magnitude for our TESS sample. Note that the contrast obtained is both larger and extends to greater $\Delta$ magnitudes at 832 nm while both filters show a convergence in overall contrast range toward fainter stars.}
  \centering
\end{figure}

\section{Detected Companions}

As an example of one of our speckle imaging reduced data products, Figure 3 presents typical contrast curves we obtain with NESSI observations at WIYN. The field of view of NESSI in speckle mode (0.018 arcsec/pixel) is 19 X 19 arcsec, however we typically only read out a 256 X 256 pixel subsection region of interest centered on the target star yielding a final image of 4.6 arcsec on a side. However, as mentioned earlier, speckle decorrelation occurs within the atmosphere outside of $\sim$1 arcsec; therefore, we only use the robust Fourier analysis and speckle reconstruction techniques, as well as determine the 5$\sigma$ contrast limits (blue points) and our fit to them (red curve) for an angular size patch of sky of 1.2 arcsec on a side, beyond which decorrelation occurs, becoming worse with increasing separation \citep{2011AJ....141...45H}.
Figure 3, showing the binary exoplanet host star TOI 894, reveals the detected close companion as a ``+" sign below the 5$\sigma$ contrast curves as well as displaying the companion star (located to the upper right) in the speckle reconstructed image insets displayed under each curve. 

\begin{figure}[th!]
\centering
  \includegraphics[scale=0.65,keepaspectratio=true]{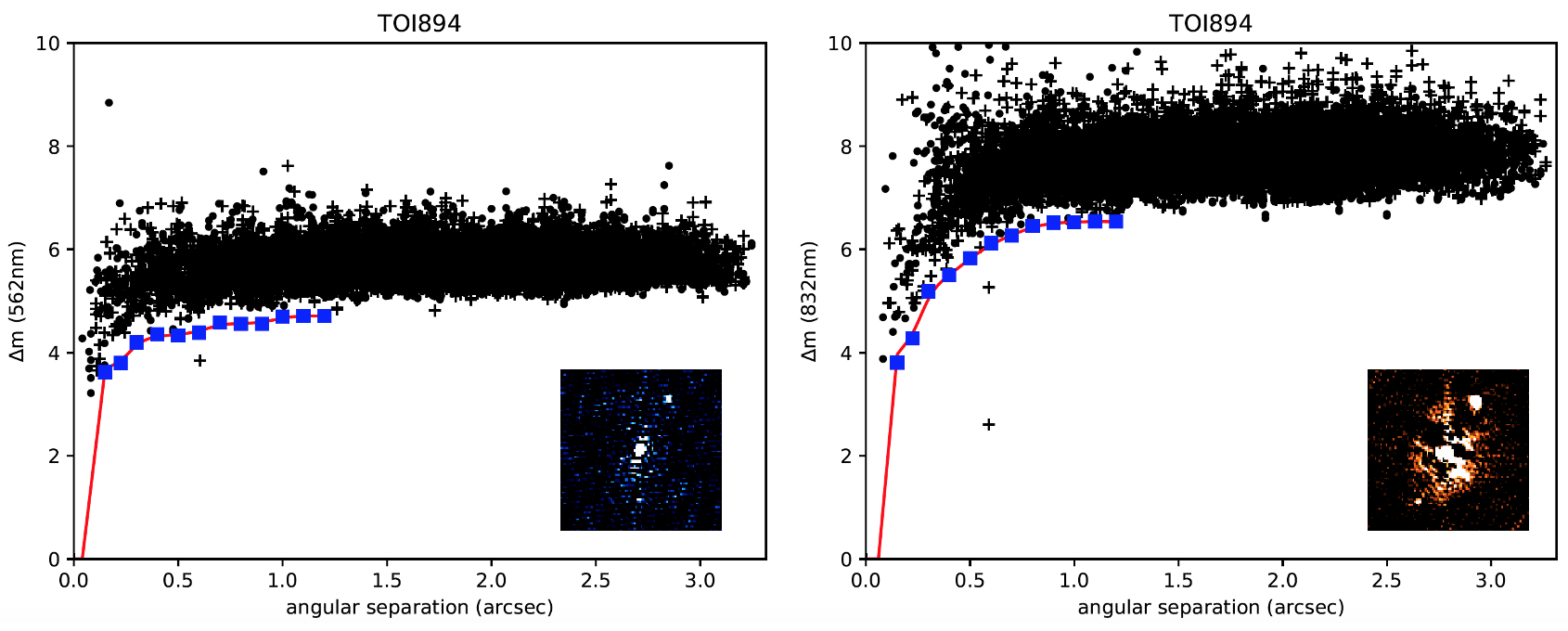}
  \caption{Speckle imaging contrast curves and reconstructed images for TOI 894. The red curve in each plot is a fit to 5$\sigma$ blue points measured at various annuli. The black + signs and filled square symbols are background measurements of the limiting contrast in the reconstructed image; + sign being points above the mean ``sky". Note the detection of the companion star at $\sim$0.5 arcsec separation, well beyond the 5$\sigma$ limit. The companion is also seen in the approximate 1 arcsec square inset images which have N up and E to the left.}
  \centering
\end{figure}

Table 2 lists the companion stars we have detected, many seen in both band-passes with fainter and/or redder companions detected only in the 832nm observation. The table gives the target name, angular separation, position angle, and $\Delta$magnitude within each respective band-pass with global internal uncertainties near $\pm$6 mas, $\pm$2 degrees, and $\pm$0.2 magnitudes, respectively. Similar values for ``Sep" and ``PA" between the two filters provide additional confirmation of the goodness of fit of the Fourier analysis. The last column presents an estimate of the orbital separation of the two stars, in au, using the distance given in Table 1 and assuming that the instantaneous spatial separation detected in our imaging is approximately the orbital semi-major axis.
Thirteen stars have companions beyond 1.25 srcsec, four of which, TOI851, TOI944, TOI994, and TOI1162, have fairly wide separated companions in the ``speckle world" (well beyond $\sim$1.25 arcsec), thus their $\Delta$mag values will not be as accurate as the rest, perhaps being overestimated and with an additional uncertainty of $\pm$0.5 magnitudes \citep{2011AJ....142...19H}. Stars with companions within $\sim$30 au will have binary orbital periods of decades and be systems that will reveal orbital motions within only a few years (See Colton et al., 2020 submitted). Two TOIs, 1163 \& 1228 are shown to be triple systems having a close companion with a wider tertiary.

\begin{figure}[h!]
\centering
  \includegraphics[scale=0.45,keepaspectratio=true]{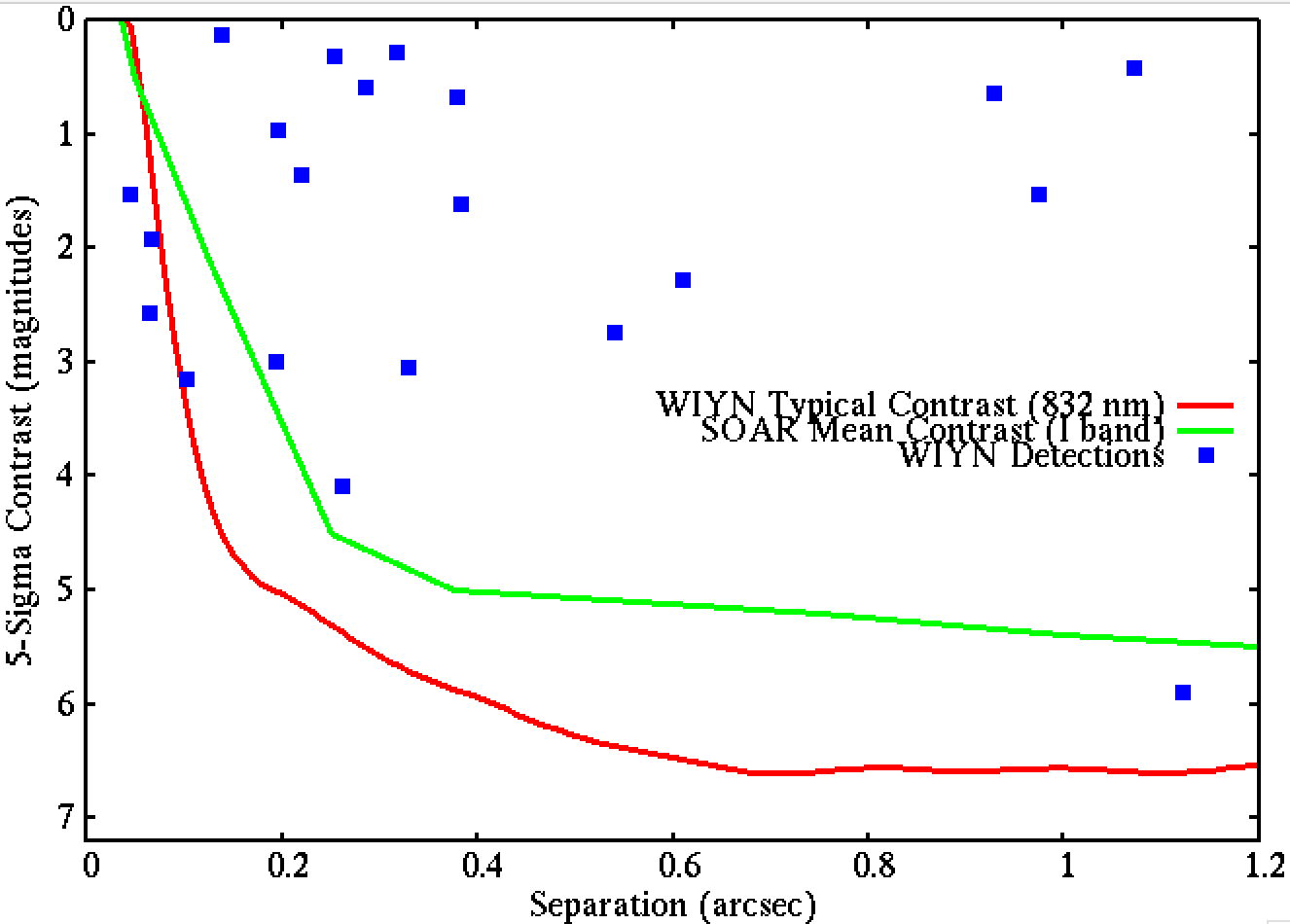}
  \caption{Speckle imaging 5-Sigma mean contrast curves. The SOAR I-band curve is from Ziegler et al. (2020) and the WIYN 832 nm curve is from Scott et al. 2018. The two contrast curves essentially match near the diffraction limit of the telescopes (SOAR is 4.1-m and WIYN is 3.5-m), 
  with the WIYN 832 nm contrast limit being deeper at all separations. 
  The stellar companions detected at WIYN inside 1.2 arcsec are shown including four inside 0.1 arcsec that would not be detected at SOAR.}\label{fig4}
  \centering
\end{figure}

\citet{2020AJ....159...19Z} recently published a paper describing TESS TOIs observed by speckle interferometry at the 4.1-m SOAR telescope in Chile. SOAR and WIYN have similar apertures, so a comparison of any common systems is useful. Not many TOIs were common between the SOAR program and ours at WIYN as TESS surveyed the southern sky in year 1 and the northern sky in year 2. However, five TOIs with detected companions (123, 172, 462, 851, 952) were observed by both telescopes/instruments and in all cases the derived parameters for the detected companions were in complete agreement.

Figure 4 compares the 5$\sigma$ contrast curves for NESSI at WIYN and the speckle imager at SOAR. The curves roughly match near the diffraction limit of these similar size telescopes while the WIYN 832-nm contrast curve limit is deeper at all separations. The companions detected at WIYN inside 1.2 arcsec (the range of good speckle interferometric correlation) are also shown in Figure 4, including four very close companions that would be undetectable at SOAR. These four exoplanet host star bound companions orbit very close to the primary star (35, 35, 16, and 7 au) and contain planetary systems with close-in (periods of 1 to 5 days), Neptune-size (radii of 5 to 9 R$_{e}$) planets.

Of the 186 TESS exoplanet host stars (TOIs) discussed in this paper, 45 total companions were found, 36 are within 1.2 arcsec. More interestingly, 21 of the companions are within 0.5 arcsec and 9 are inside 0.25 arcsec - companions difficult to impossible to detect by other means. 
Our speckle imaging program is aimed at the detection of companions that can cause validation and exoplanet characterization problems and can cause spectral modeling to produce incorrect values for the host star properties \citep{2020ApJ...898...47F}. We are particularly interested in finding true, bound companions, stars that generally lie inside 0.25 arcsec \citep{2018AJ....156...31M}, which are hard or impossible to discover with other instruments or other means.
These true bound companions are the stars which will provide detailed robust information for exoplanet formation, dynamics, and evolution. Finally, our simultaneous two-color observations allow us to not only detect companions, but also gain some astrophysical knowledge for them in terms of the companion spectral type (i.e., mass, see below).

\section {Discussion}

Speckle interferometry has become one of the leading ground-based telescope follow-up resources in the study of exoplanets. Such observations are critical to obtaining correct stellar and exoplanet properties. While both the Kepler and K2 mission exoplanet host stars yielded a similar percentage of binaries (near 40-50\%; \citet{2014ApJ...795...60H}; \citealt{2016MNRAS.455.4212D}; \citealt{2018AJ....156...31M}; \citealt{2018AJ....156..259Z}), the missions sampled different regions of the sky (single field of view at mid-galactic latitude vs.~multiple ecliptic fields of view) as well as having different observing strategies and photometric precision levels. 

The TESS survey affords us with an opportunity to directly compare the observed stellar companion detection fractions and binary host star properties for nearly equal surveys in the southern and northern sky. We acknowledge that the speckle interferometric observed companion stars have a bias (as is true of any observational campaign) and do not not represent the full story. The companions that are missing could be those with large magnitude differences (below the contrasts available), orbital locations which place them inside even the resolution of speckle imaging (very close binaries), companions of very red color (very faint in the optical band-pass), and companions at separations outside the field of view (i.e., outside the angular distance for speckle correlation, near 1.2"). In our speckle imaging program, we concentrate on nearby companions ($<$1.2 arcsec), stars with a high probability of being true bound companions and essentially undetectable by other means. In our analysis, we account for possible ``missing" companions in a statistically probabilistic manner (see below).  

Using $<$1.2 arcsec as a guide, we find that 24 out of our total of 186 stars show a detected companion, giving a detection fraction of 13\%. \cite{2020AJ....159...19Z} find 67 stars (out of 542 total stars they observed) with detected companions within 1 arcsec or 12.5\%. This is gratifying to see as it lends credence to the fact that the exoplanet host stars which are binaries seem to yield the same observed percentage in both hemispheres when using comparable (but different) instruments and telescopes.
It is interesting to note that the observed percentage of 13\% is roughly twice that seen for Kepler and K2 exoplanet host stars observed with speckle imaging at the WIYN telescope, 7\% \citep{2014ApJ...795...60H} and 6\%
\citep{2018AJ....156...31M}, but are not unexpected based on the observed 46\% true binary fraction and the generally much closer and brighter stars being observed by  TESS. See \citet{2019AJ....157..211M} and \S4.1 below. 

\begin{figure}[h!]
\centering
  \includegraphics[scale=0.45,keepaspectratio=true]{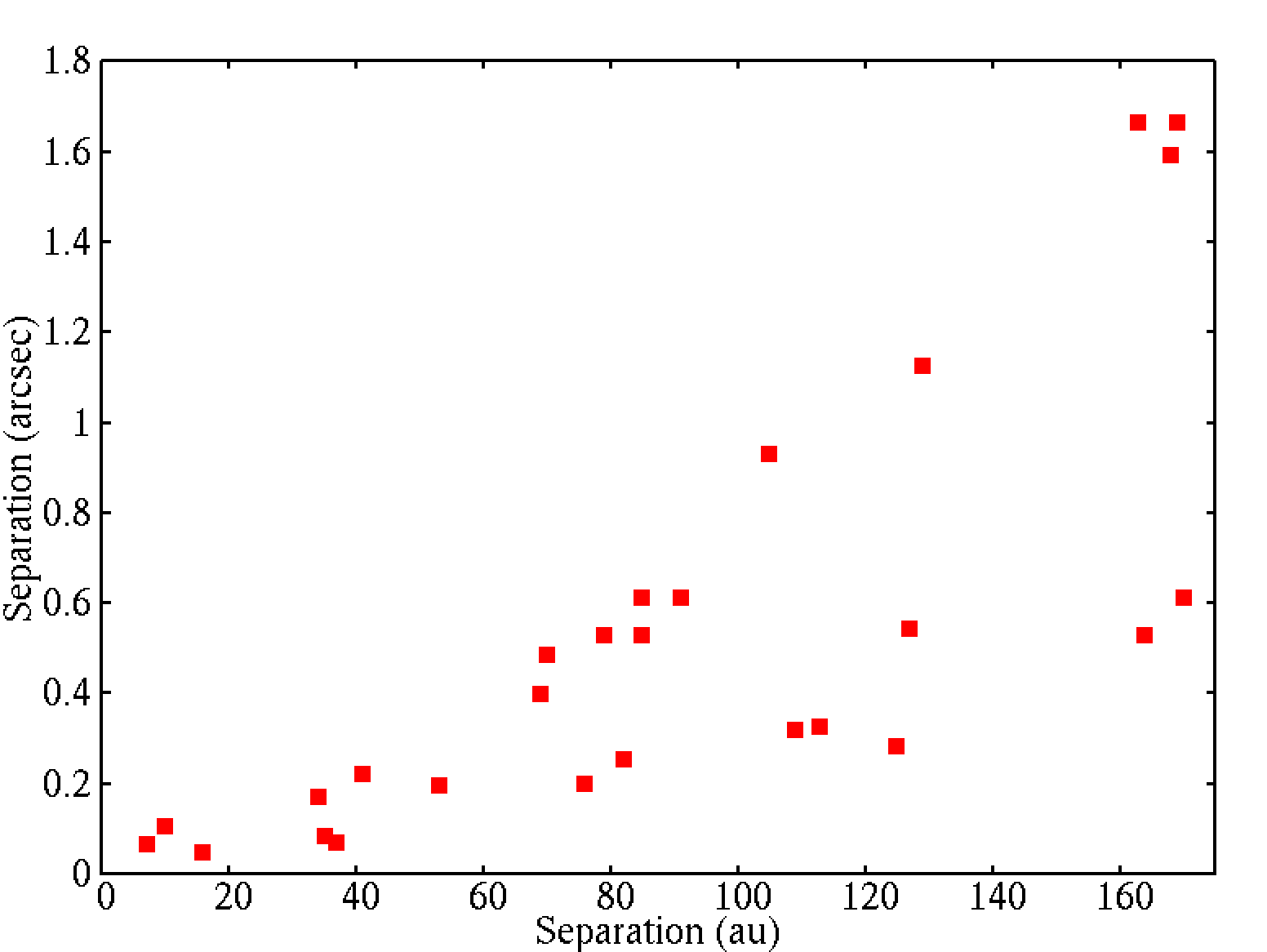}
  \caption{The relationship between angular separation (arcsec) and spatial separation (au) of the detected companions within a projected distance of 200 au (1.6"). We note that the closest spatial separations make up the majority of the closest physical separations. There is a dropoff in the frequency of detected companions outside of $\sim$0.6 arcsec, possibly representing the transition from mostly bound to mostly line-of-sight companions.  \label{fig5}
  }
  \centering
\end{figure}





Figure 5 shows the relation between the angular separation in arcsec of our detected companions and their projected physical separations in au. We can see in the figure the existence of a near-linear cutoff line for the minimum spatial separation observable as a function of the angular separation available for the TESS targets. Outside of 0.6 arcsec, we see a falloff in the number of detected companions perhaps signaling a true drop-off in bound companions vs.~line-of-sight companions (See \citealt{2018AJ....156...31M}). Of course, gravitationally bound systems do indeed exist at larger separations than 0.6 arcsec (e.g., Common proper motion pairs), however they are increasingly rare to detect as the separation becomes larger than our field of regard.
Using Figure 5, we note that for a typical TESS star ($<$500 pc), the contrast curves in Figure 4 begin to turn over at their inner working angle at values of $\sim$46 au (0.2") at SOAR and $\sim$14 au (0.1") with NESSI at WIYN.


\subsection{Predictions of Stellar Companions}

\begin{figure}[h!] 
\centering
  \includegraphics[scale=0.9,keepaspectratio=true]{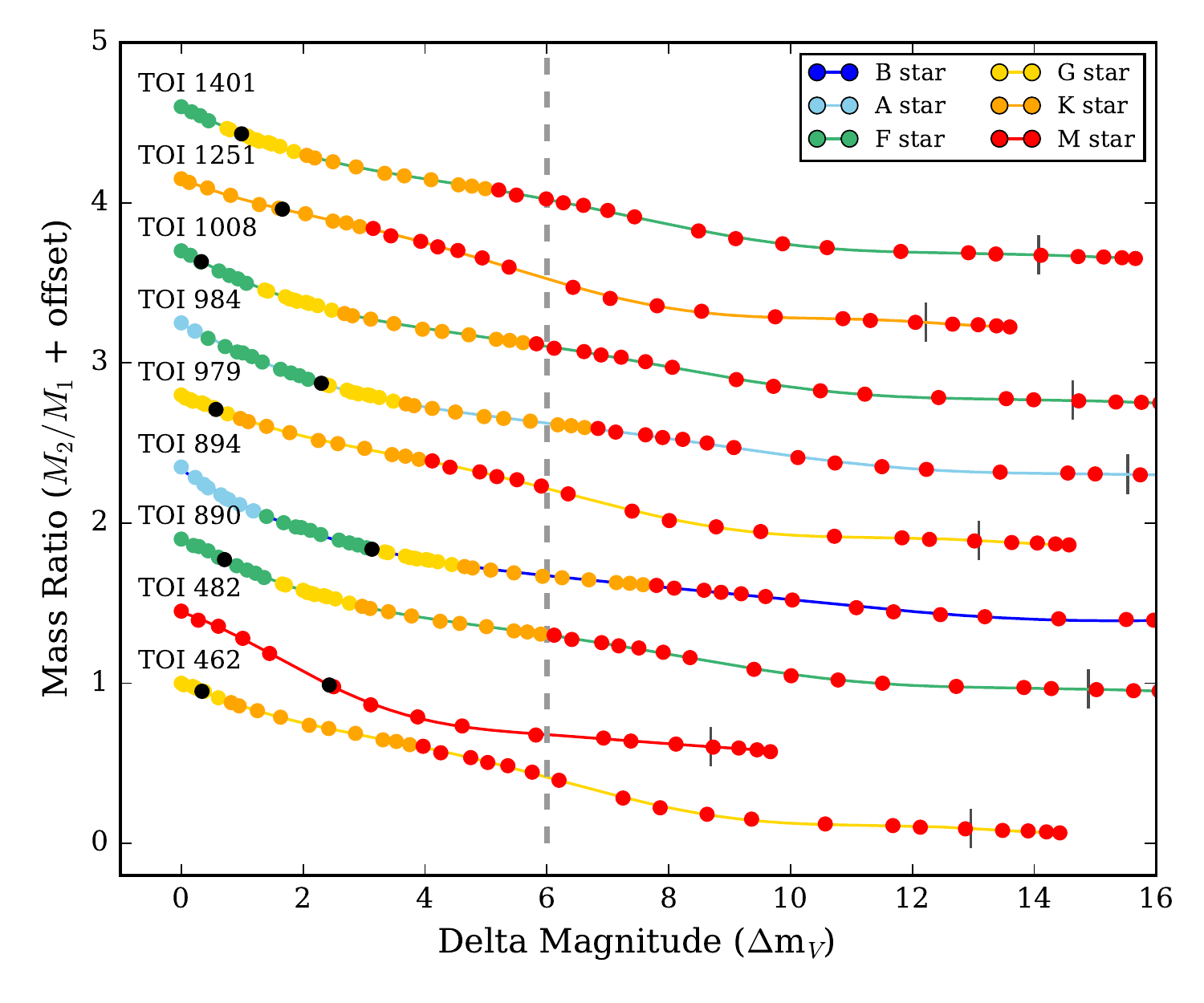}
  \caption{Expected mass ratio (offset for clarity) vs.~$V$-band delta magnitude for a sample of TOIs with detected companions. The colors of the lines and dots correspond to the spectral type of the star/possible companion. Contrast limits for speckle imaging at 562nm with NESSI are shown by the dashed line ($\Delta m \lesssim 6$). The small vertical lines along each line indicate where the mass ratio equals 0.1. The black dots show the measured delta magnitude (562nm, except 832 nm used for TOI 482) for observed companions. See the Appendix for a plot of all detected companions. }\label{fig:compfrac}
  \centering
\end{figure}

\citet{2019AJ....157..211M} used predicted distributions of TESS exoplanet host stars to examine the population of stellar companions detectable with speckle imaging. Here we use the same techniques to compare our observed companions to expectations and estimate the stellar parameters for the companion stars. We begin by considering all possible bound companions for a given TOI using the Modern Mean Dwarf Stellar Color and Effective Temperature Sequence Table based on \citet{2013ApJS..208....9P} to assign stars with later spectral types (cooler effective temperatures) as possible companions. For each possible companion we then calculate a $V-$band delta magnitude ($\Delta$m) and mass ratio ($q = M_{2}/M_{1}$) relative to the primary (TOI) star. Because we assign companions based on spectral types, which are discrete and unevenly spaced in mass, we fit a seventh order polynomial to the mass ratios as a function of delta magnitudes for each TOI and then determine the fraction of companions that fall within the 562 nm speckle contrast limit of $\Delta$m $\leq 6$. We also weight the likelihood of each companion by the mass ratio distribution of \citet{2010ApJS..190....1R}, such that companions with mass ratios of $0.1 \leq q \leq 0.95$ are equally likely and those with q $>$ 0.95 are enhanced by 2.5$\times$. Figure \ref{fig:compfrac} shows possible companions for select TOIs, with the spectral type of the TOI indicated by the solid line color and the spectral types of the individual possible companions indicated by the color of the points. The dashed line highlights the speckle detection limit at $\Delta$m = 6, and small vertical lines indicate where the mass ratio is equal to 0.1 for each TOI. The fractions of detectable companions for all TOIs in our sample are listed in Table~\ref{tab:comptable} under “Comp. Frac.”, with a mean of 0.57 for the sample. Table \ref{tab:comptable} also shows the stellar parameters used in our predictions as well as the spectral type of potential companions at $\Delta$m = 4 and $\Delta$m = 6 for all TOIs observed at WIYN. Two stars without distances are listed for completeness, but no analysis of possible companions is included.

\begin{figure}[h!] 
\centering
  \includegraphics[scale=0.4,keepaspectratio=true]{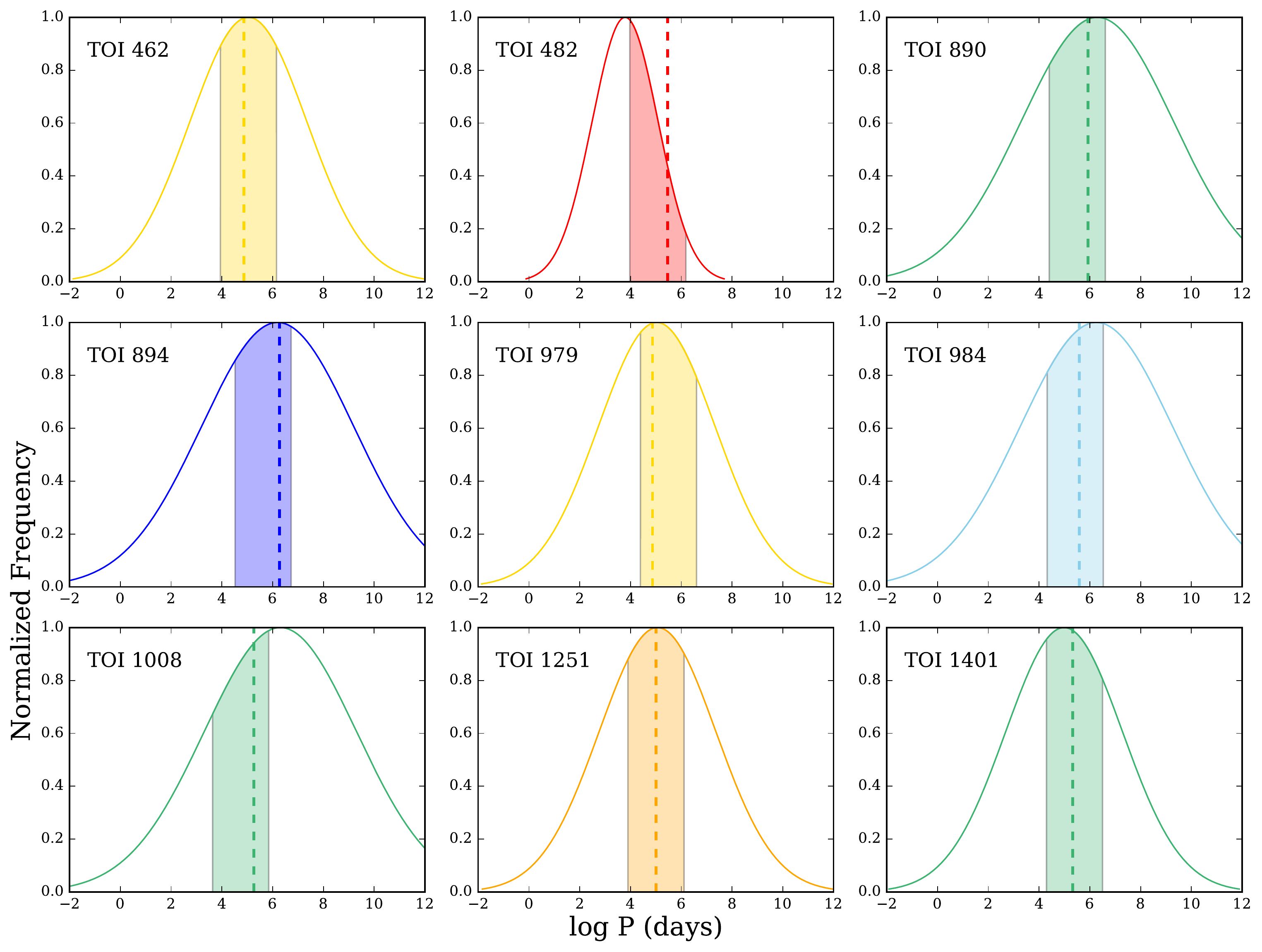}
  \caption{Expected binary period distributions for a selection of TOIs with detected companions based on the distribution presented in \citet{2010ApJS..190....1R}. The shaded regions (color coded by spectral type of the primary as in Figure \ref{fig:compfrac}) show the orbital periods corresponding to projected separations at which speckle imaging at WIYN (562nm) can detect companions. The dashed lines correspond to the separation of observed companions converted to $\log P$ space using the distance and mass of the primary star. See the Appendix for a plot of all binary period distributions.}\label{fig:distrfrac}
  \centering
\end{figure}

Since the separation of the components must also be considered when evaluating which companions speckle imaging is sensitive to, we use the binary orbital period distribution for solar-type stars \citep{2010ApJS..190....1R} to determine the fraction of possible companion separations that can be detected. For each TOI we use a log-normal period distribution parameterized by the mean semi-major axis in au and standard deviation in $\log P$ based on the mass of the primary (See Table 3 of \citealt{2015ApJ...809...77S}). The angular resolution limits of NESSI (0.04 - 1.2" at 562nm) are then converted to period space using the mass and distance of each TOI via Kepler's third law. The binary period distributions and portions of $\log P$ space that are detectable using speckle imaging for TOIs with a range of spectral types are shown in Figure \ref{fig:distrfrac} (shaded regions). The speckle limits in au and the fraction of each binary distribution, ``Distr.~Frac.'', that falls within those limits (converted to $\log P$) are listed in Table \ref{tab:comptable}. The mean 
``Distr.~Frac.'' for all TOIs observed at WIYN is 0.34. Taking the mean value for ``Distr.~Frac." times the mean value for ``Comp.~Frac." (0.57 from above), we should expect to actually observe bound companions around $\sim$19\% of our stars, a value comparable to our observed value of 13\%. 

Figures \ref{fig:compfrac} and \ref{fig:distrfrac} also show where observed companions from this study were detected in relation to our predictions. As in the predictions, we assigned the spectral type of the primary star based on the $T_\mathrm{eff}$ reported in Table 1, and then determine the spectral type of the secondary using the observed delta magnitude. For systems with small separations and small delta magnitudes, such as TOI 890, 979, 1008, and 1267, the reported effective temperature will be a combination of both stars. While disentangling the temperatures of the two components in these systems does not impact our overall results, it should be considered for detailed analysis of the individual systems. The 562 nm delta magnitude of companions observed around the TOIs in Figure \ref{fig:compfrac} are shown as black dots (832 nm shown for TOI~482), while the observed separations for the same companions are indicated by dashed lines in Figure \ref{fig:distrfrac}. Plots showing our predictions and the observed delta magnitudes and separations for all TOIs with detected companions are given in the Appendix (Figures~11-14).

Similar to Figure \ref{fig4}, the plot of delta magnitude vs.~mass ratio illustrates that most companions detected at WIYN have $\Delta$mag $\leq 3-4$ and have companions of the same or slightly cooler spectral types. We estimate the mass ratio for each observed system by converting the predicted $V-$band delta magnitude to a TESS delta magnitude (via \citealt{2018AJ....156..102S}) and fitting a polynomial to the possible mass ratios as a function of $\Delta$m$_{TESS}$. We then use the observed 832nm $\Delta$m values for each TOI (since all companions were detected in that filter) to determine the mass ratio from the polynomial fit. 

\begin{figure}
\gridline{\fig{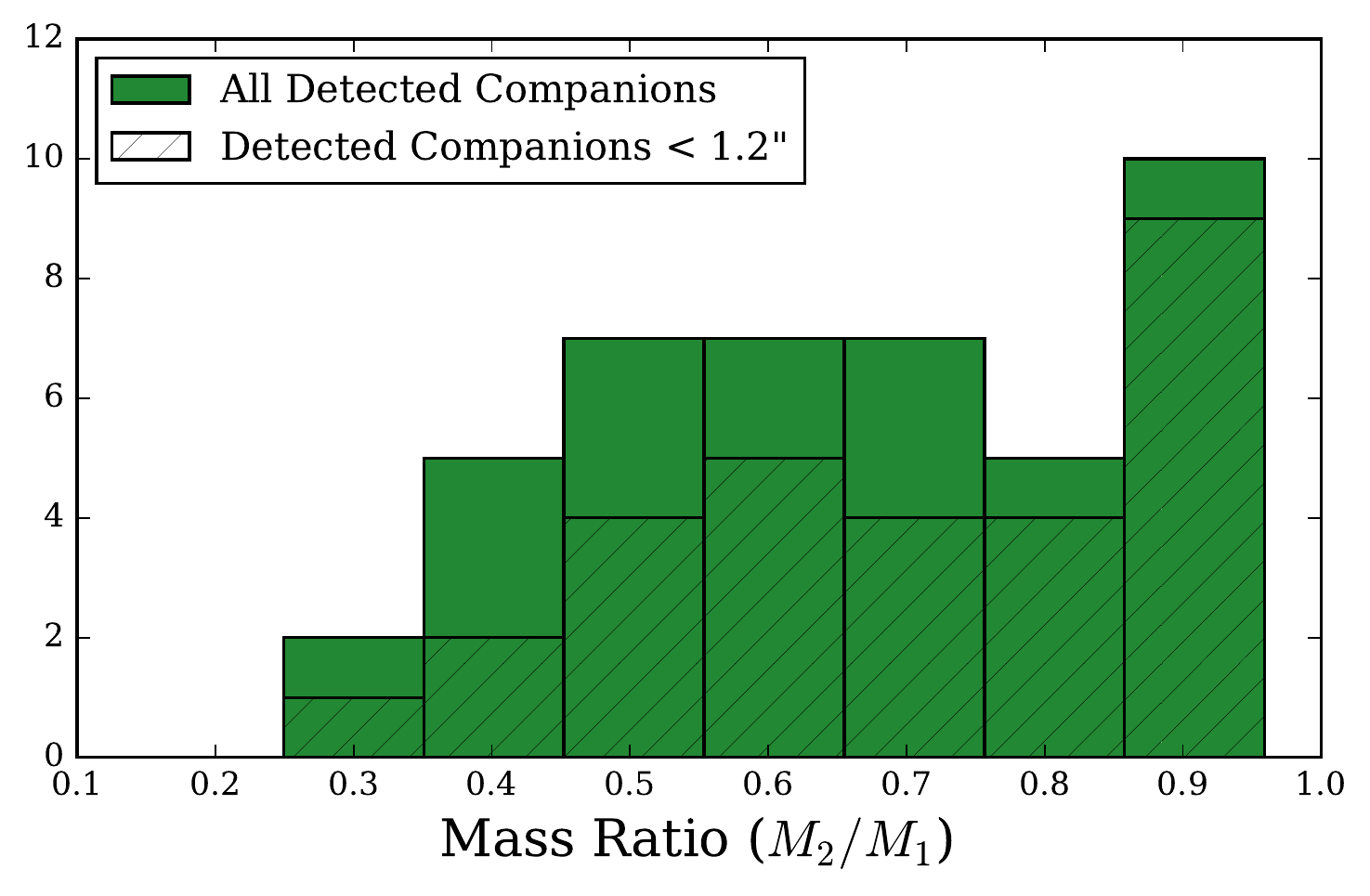}{0.45\textwidth}{}
          \fig{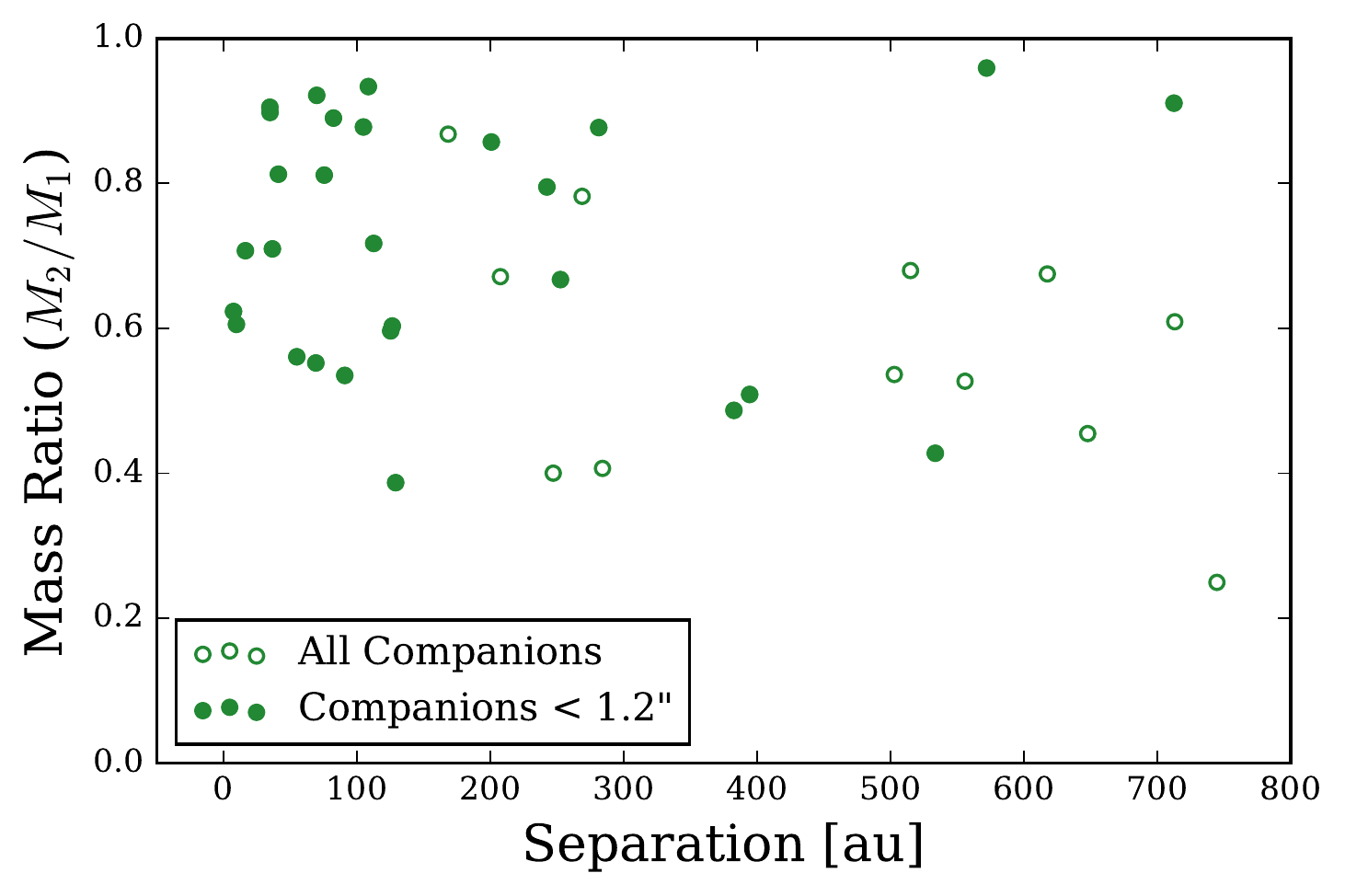}{0.45\textwidth}{}}
\caption{Left: Calculated mass ratios for all companions detected with speckle imaging at WIYN including those detected within 1.2" (hatched bars). The mass ratios were determined using the measured $\Delta$mag at 832nm and a polynomial fit to the mass ratio as a function of TESS delta magnitude based on \citet{2013ApJS..208....9P}. See Figure \ref{fig:compfrac} and the text for more details. Right: Mass ratio as a function of separation in au for all companions (open points) and those detected within 1.2" (solid points).}
\label{fig:qhist}
\end{figure}

A histogram of the mass ratios for all detected companions is shown in Figure \ref{fig:qhist}, with the companions detected within 1.2" shown as hatched bars. The mass ratio distribution shows an increase in the number of high-q systems with a uniform distribution at lower-q, in agreement with \citet{2010ApJS..190....1R}. The drop off at the lowest q-values ($\leq$0.4) is where our sensitivity to faint, red companions decreases. Figure \ref{fig:qhist} also shows the mass ratios as a function of separation in au, with the companions detected within 1.2" shown as solid points. While the lack of low-q systems at smaller separations could be taken as an observational bias (and likely is), the relative lack of close ($<$1.2") high-q systems at larger separations is indeed interesting and suggestive. 
Similarly, \citet{2010ApJS..190....1R} opined that short-period systems prefer higher mass ratios, that is,
like-mass pairs (q$>$0.95) prefer relatively short orbital periods. Likewise, \citet{2017ApJS..230...15M} noted that the ``twin fraction" significantly decreases with orbital period.
We also see a slight preference for the hotter stars to be in high-q systems whereas the cooler stars tend toward low-q systems. 

For the stars shown in Figure 8, we examined in detail possible connections of planet radius or orbital period with the stellar properties. Neither exoplanet orbital period (ranging from $<$1 to 15 days, well within even our closest binary systems at $<$10 au) nor exoplanet radius ($<$1 to 30 R$_e$) showed a meaningful correlation with the stellar or binary properties in this relatively small sample. Nine of the 25 exoplanets in these binaries are larger than 10R$_e$ and generally belong to wider separated pairs ($>$200 au). TOI 1191\footnote{TOI 1191 is also probably a triple star system, with a second companion recently detected at a separation of 0.7 arcsec and 5 magnitudes fainter (Lester et al., in prep.).} is a notable exception harboring a planet of radius 25.4R$_e$, orbital period of 15.7 days, and residing in a binary system separated by only 16 au.

Figure \ref{fig:distrfrac} confirms that observed companions also fall within our expected separation/period ranges and highlights our sensitivity to the peak of the expected binary period distribution for most TOIs, making speckle imaging ideal for companion searches. While further information is needed to robustly prove whether the observed companions are truly bound, the fact that the delta magnitudes and separations we measure correspond to realistic binary parameters gives us confidence that most close companions are likely to be physically bound.

\begin{figure}[ht] 
\centering
  \includegraphics[scale=0.65,keepaspectratio=true]{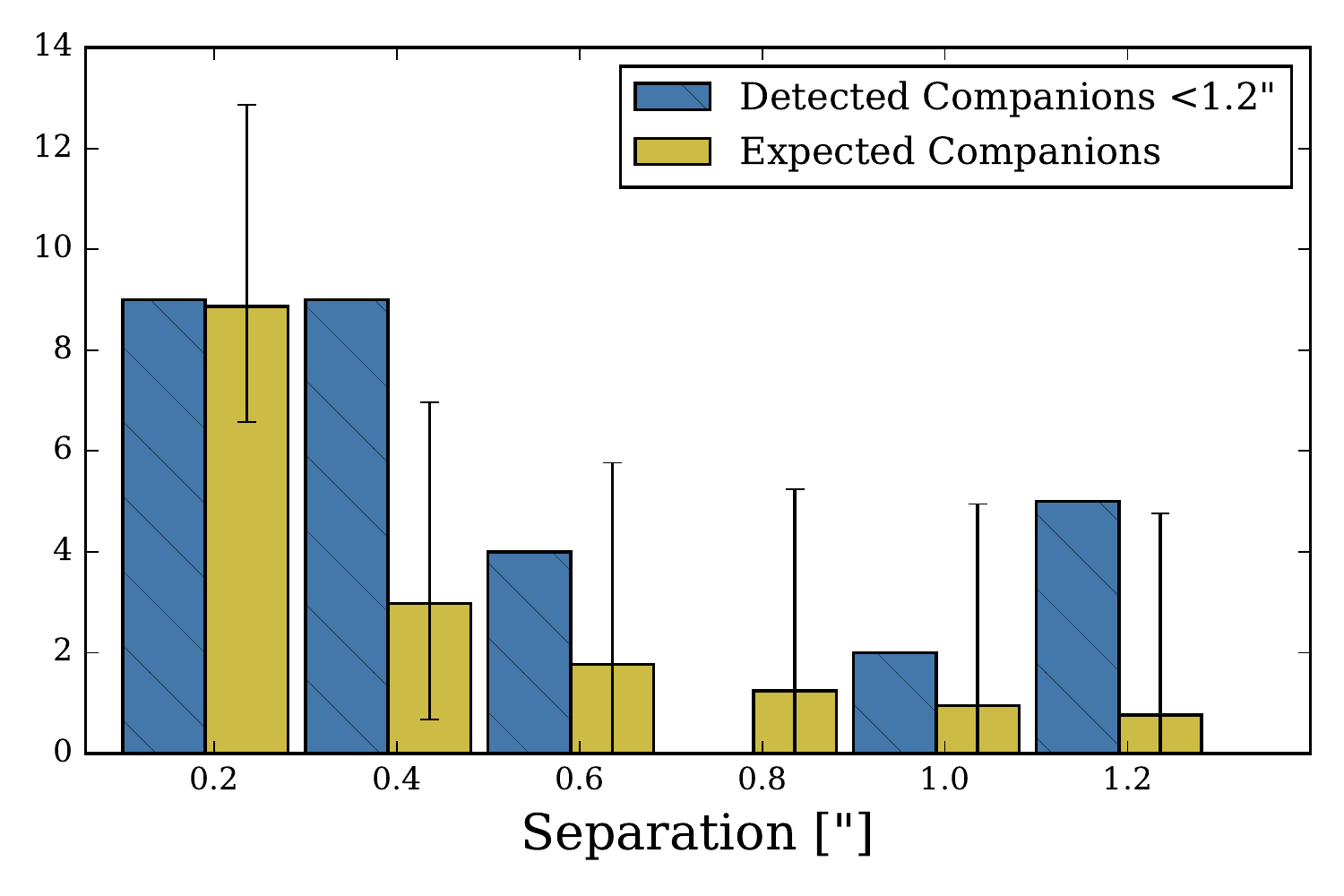}
  \caption{Histogram of companion stars observed within 1.2" of a TOI (hatched blue bars) and the number of expected companions (solid yellow bars) in projected separation bins of 0.2". The number of expected companions was derived from the fraction of companions that can be detected around 46\% of the TOIs using speckle imaging at WIYN. Error bars on the expected companions were determined from the minimum and maximum number of expected companions based on 100000 random combinations of 46\% of the TOIs in our sample. See text for more details.}\label{fig:sephist}
  \centering
\end{figure}

We also use our fraction of possible companions (``Comp.~Frac.'') and fraction of each binary period distribution (``Distr.~Frac.'') that NESSI can detect to estimate the number of companions we expect to observe around the 186 TOIs observed at WIYN. For each TOI we determine the ``Comp.~Frac.'' and ``Distr.~Frac.'' in separation bins of width=0.01" from 0.04 to 0.24" and bins of width=0.2" from 0.24 to 1.24", to account for the rapidly changing delta magnitude detection limits at close separations (see Fig.~\ref{fig4}). We then multiply the ``Comp.~Frac.'' by the ``Distr.~Frac.'' for each TOI (see ``Speckle Frac.'' in Table \ref{tab:comptable}), and sum the fraction of companions in each separation bin to determine the total number of companions we expect to detect with speckle imaging. Assuming 46\% of the TOIs have true bound companions, we multiply the total number of detectable companions by 0.46 to get $\sim 17$ expected companions within 1.2". The results are plotted in Figure \ref{fig:sephist}, with the expected number of companions shown in separation bins of 0.2" as yellow bars, and the observed companions within 1.2" plotted as blue-hatched bars. The error bars on the expected companions reflect the minimum and maximum total number of expected companions determined by randomly selecting which 46\% of the TOIs have companions for 100,000 random iterations. While our calculations only expect $\sim17$ companions compared to the 29 we observe within 1.2" of 28 TOIs, the number of companions observed matches the model predictions given the sizable uncertainties in each bin (See Figure \ref{fig:sephist}). 

\begin{figure}[t!] 
\centering
  \includegraphics[scale=0.8,keepaspectratio=true]{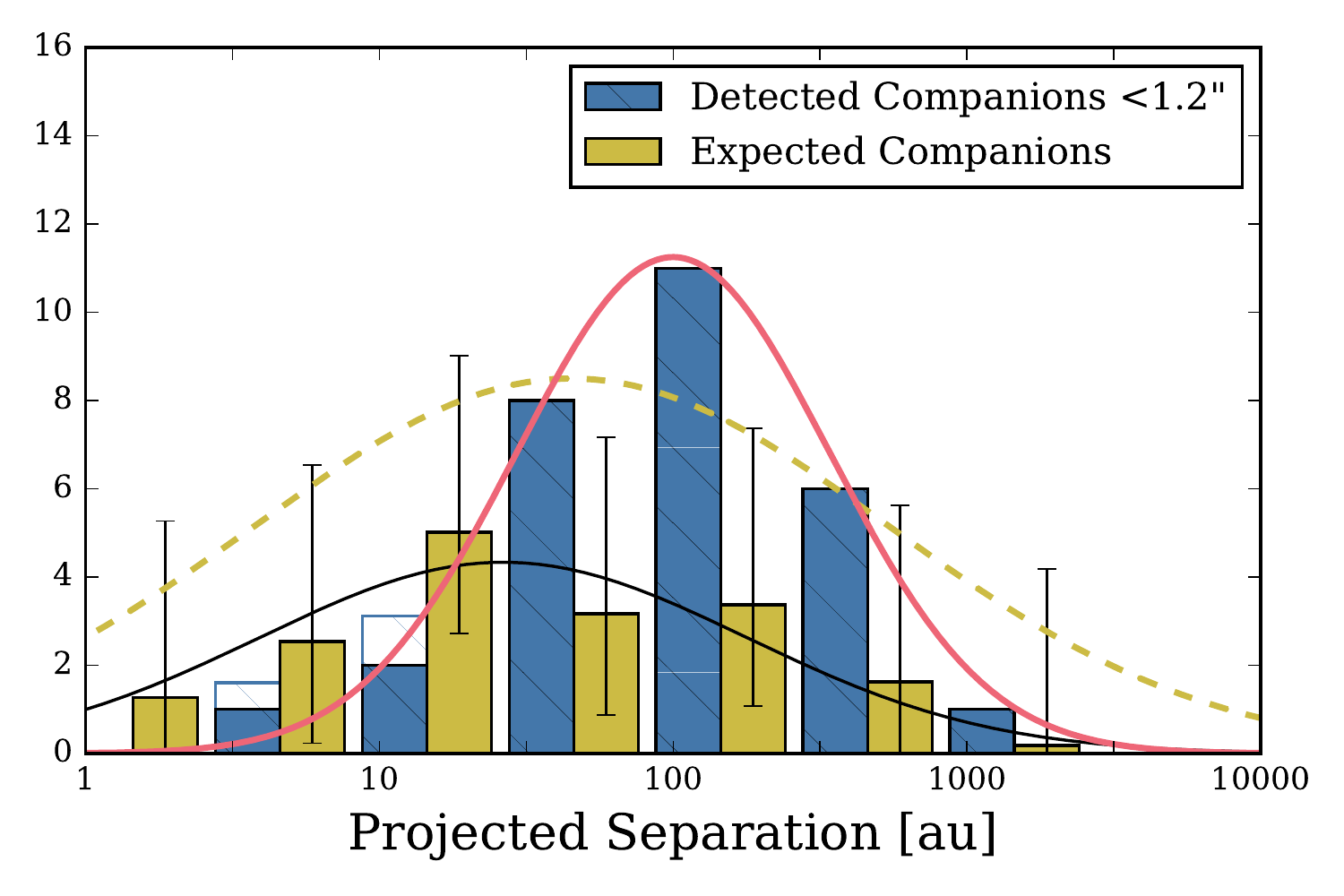}
  \caption{Histogram of projected separation in logarithmic bins of au for companions detected within 1.2" of a TOI at WIYN (solid blue bars). The unfilled hatched bars represent a `correction' factor based on our mean contrast curve range in delta magnitude over the separation bins in au (see Figure 4 and text for more details). A Gaussian fit to the `corrected' companion distributions is shown in red, which is narrower and peaks further out than the distribution of \citet{2010ApJS..190....1R}. Also plotted is the expected number of companions (solid yellow bars) as determined from the ``Comp.~Frac.'' and ``Distr.~Frac'' of 46\% of the TOIs (see text for details). The Gaussian fit to the expected companion distribution is shown in black, while the Raghavan distribution, scaled to the expected number of companions and bin width, is plotted as a yellow dashed line. }\label{fig:sepauhist}
  \centering
\end{figure}

To further examine the population of observed companions and determine whether we see any indication of binary suppression (inside of $\sim$40 au) among close binaries with exoplanets as suggested by \citet{2016AJ....152....8K} and \citet{2020AJ....159...19Z}, we convert the projected separation in arcseconds to au using the distance of each TOI for both the observed and expected companions. Figure \ref{fig:sepauhist} shows the separation in au of observed companions within 1.2" in logarithmic bins of 0.5 dex (hatched blue bars).

However, the two left most bins are underrepresented in our sample not due to binary suppression, but due to the speckle contrast curve becoming steep and shallow at small separations. This is a property of every high-resolution imaging instrument, with the inner working angle efficiency falling off at $\sim$14 au at WIYN  and $\sim$40 au at SOAR. Given this increased incompleteness at small separations (See Figure 4) we estimate a ``correction" factor based on a convolution of the inner contrast curve, that is the steep change in magnitude averaged over the corresponding broad bins in au (taking min au vs.~separation as a metric for minimum separations observable at each bin of arcsec, see Fig.~\ref{fig5}). This correction increases the left most bins by 0.6 and 2.1 stars, respectively (unfilled hatched bars in Fig.~\ref{fig:sepauhist}). Fitting a Gaussian to the corrected distribution (solid red line) shows that the TESS companion sample peaks at $\sim 100$ au with a width of 0.75 au. {\textbf{The Gaussian distribution was determined using a non-linear least squares fit to the ``corrected'' bins, but is consistent with the fit determined from a maximum likelihood estimate on the unbinned data.}}

Figure \ref{fig:sepauhist} also shows the expected number of companions (solid yellow bars) based on the Raghavan distribution convolved with the NESSI contrast curve as described previously. A Gaussian fit to the expected companions, {\textbf{using a non-linear least squares fit to the binned data}}, (solid black line) peaks at a separation of only 26 au, a value biased toward close separations given the NESSI contrast curve convolution with the distribution of \citet[][peak $\sim$ 50 au, $\sigma$ $\sim$ 1.1]{2010ApJS..190....1R}. The uncertainties on the expected companion bins, however, allow the distribution to be statistically consistent with the Raghavan distribution, shown in Figure \ref{fig:sepauhist} as a dashed yellow line, which has been scaled to the expected number of companions (17) and bin widths.

\begin{figure}[t!] 
\centering
  \includegraphics[scale=0.8,keepaspectratio=true]{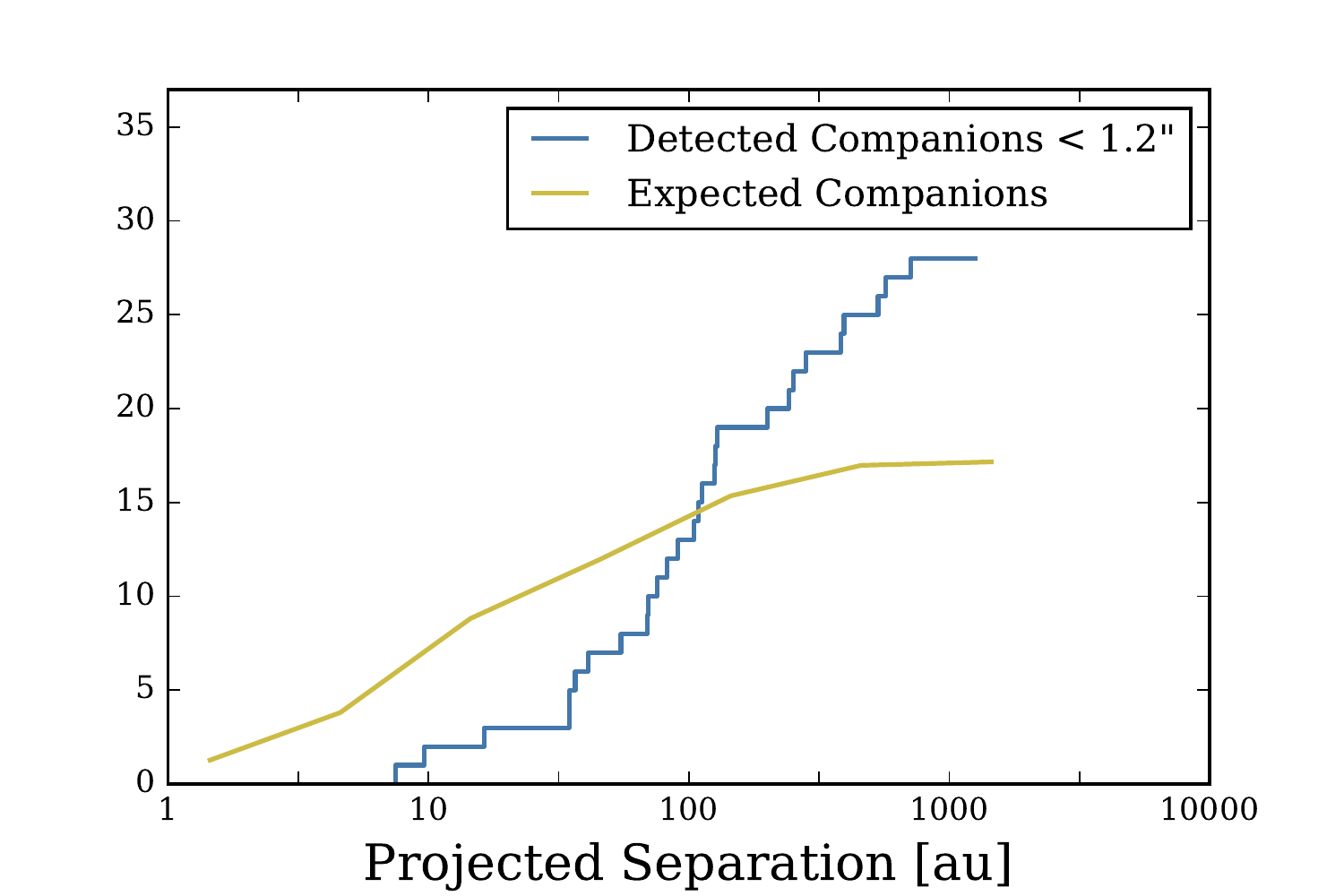}
  \caption{Cumulative distribution for projected separations of companions detected within 1.2" of exoplanet candidate host stars using speckle imaging at WIYN. The expected companion distribution based on the Raghavan distribution convolved with the NESSI contrast curve (see text for details) is shown for comparison. The expected companion distribution is in logarithmic bins of 0.5 dex as in Figure~\ref{fig:sepauhist}. The Y-axis is the number of companions. \label{fig:cumul}}
  \centering
\end{figure}

\textbf{We also present the unbinned cumulative distribution of detected companions within 1.2 arcsec in Figure~\ref{fig:cumul}. The (binned) expected companion distribution from Figure~\ref{fig:sepauhist} is shown for comparison.}

Based on Figures \ref{fig:sepauhist} and \ref{fig:cumul}, we note that our distribution of observed companion separations does not match that of \citet{2010ApJS..190....1R}. While we have only a modest sample of binary exoplanet host star companions, there is evidence in Figures 8-10 that binary stars which host exoplanets have a ``Raghavan-like" mass distribution but have a larger mean orbital separation ($\sim$100 au) with a slightly narrower breadth in their distribution. 
Companion stars can eject planets from stellar systems \citep{2006ApJ...644..543H}, perturb proto-planetary disks
\citep{2015ApJ...799..147J}, and create planetesimals through gravitational interactions \citep{2015ApJ...798...69R}. The wider mean separation of exoplanet hosting binaries is likely needed to permit proto-planetary disk formation and planetary evolution without disruption by the stellar companion. 

Thus, discussions of close binary star ``suppression" might indeed be real, but not because of a fall-off of systems within the Raghavan distribution at close separation period, but rather due to exoplanet host star binaries having a different orbital period distribution all together, one with a larger mean separation. 




\section{Conclusion}
We have presented the first year of high-resolution speckle imaging observations for TESS stars using the NN-EXPLORE NESSI instrument at the 3.5-m WIYN telescope.
Speckle observations of 186 TESS exoplanet host stars were carried out, the majority of which were brighter than 12th magnitude, closer than 500 pc, and solar-like. Of the TESS stars observed, 
45 (13\%) revealed a close companion. This number is consistent with the expected 19\% value based on our models. The distribution of stellar mass ratios in exoplanet binary star systems seems to match that of the Raghavan field binary distribution, perhaps presenting some common aspect of binary star formation, while there may be a lack of high-q binaries at wider separations. Our measured orbital separation distribution, however, does not match the expectation. Our observations support the fact that exoplanet hosting binary stars have, in general, wider separations (longer orbital periods) than field binaries. Thus, the suggested close binary suppression is really not  ``missing" short period systems from a normal Raghavan-like distribution, but instead is a feature of a wider separation population of binaries, those that host exoplanets. 

Additional observations of binary exoplanet host stars are needed in order to build up the sample. A control sample, that is a sample of similar TESS stars not seen to show a transit event, are also needed in order to help vet observational biases. Higher contrast, higher resolution speckle imaging observations will be presented in an upcoming study of TESS exoplanet host stars using 2 years of speckle interferometric observations from the 8-m Gemini North and Gemini-South telescopes.

\bigskip
\bigskip

Note added in Proof: After our analysis was finished and this paper submitted, Gaia EDR3 was released, providing refined parallax values. These improved values, essentially all within $\pm$10\% for our stars with companions of 1.2 arcsec or less, produce only small bimodel changes to the distances and are inconsequential in relation to the main results presented in this paper.

\acknowledgments
The observations in the paper made use of the NN-EXPLORE Exoplanet and Stellar Speckle Imager (NESSI). NESSI was funded by the NASA Exoplanet Exploration Program and the NASA Ames Research Center. NESSI was built at the Ames Research Center by Steve B. Howell, Nic Scott, Elliott P. Horch, and Emmett Quigley. 
This research has made use of the NASA Exoplanet Archive and ExoFOP, which are operated by the California Institute of Technology, under contract with the National Aeronautics and Space Administration under the Exoplanet Exploration Program.
J.N.W.\ acknowledges funding from the NASA-WIYN Data Analysis program (JPL contract
1597372).

{\it {Facilities:}} WIYN -  NESSI, DSSI

{\it {Software:}} astropy (The Astropy Collaboration 2013, 2018), SciPy (Jones et al. 2001)



\startlongtable
\begin{deluxetable}{lcccccccc}
\tablecaption{TESS Stars Observed by NESSI at WIYN}
\tablehead{
\colhead{Target} & \colhead{Gaia Mag} & \colhead{T$_{eff}$} & \colhead{Dist} & \colhead{UT Date} & \multicolumn{2}{c}{$\Delta$Mag 562 nm}  & \multicolumn{2}{c}{$\Delta$Mag 832 nm} \\ 
\colhead{} & \colhead{} & \colhead{K} & \colhead{(pc)} & \colhead{MM/DD/YY} & \colhead{0.2"} & \colhead{1.0"} & \colhead{0.2"} & \colhead{1.0"}
}
\startdata
TOI 103$^{a}$ & 11.9 & 6371 & 411.2 & 06/18/19 & 3.41 & 3.77 & 3.82 & 4.68 \\
TOI 109$^{b}$&13.8&5361&513.0&06/19/19&3.55&3.72&3.75&4.46 \\
TOI 123&8.2&6356&161.5&06/19/19&3.60&4.20&4.52&6.43 \\
TOI 172&11.2&5759&342.8&06/19/19&3.41&4.04&3.83&5.44 \\
TOI 254 & 10.3 & 6101 & 133.3 & 01/21/19 & 2.81 & 3.98 & 4.03 & 6.52 \\
TOI 260 & 9.3 & 4049 & 20.2 & 01/21/19 & 3.90 & 4.82 & 4.02 & 6.60 \\
TOI 266 & 9.9 & 5784 & 101.7 & 01/21/19 & 3.51 & 4.62 & 3.66 & 6.25 \\
TOI 278&14.7&2950&44.4&10/13/19&3.46&4.23&3.47&4.77 \\
TOI 309$^c$ & 12.9 & 5329 & 345.3 & 01/24/19 & 4.10 & 4.66 & 3.85 & 4.69 \\
TOI 316&13.9&4245&275.0&10/12/19&3.59&4.09&4.04&4.80 \\
TOI 329 & 11.2 & 5560 & 284.4 & 01/23/19 & 3.09 & 3.48 & 3.64 & 4.43 \\
TOI 390&10.2&6321&167.3&10/12/19&3.19&4.35&3.19&4.35 \\
TOI 438&10.0&5211&72.5&10/10/19&4.40&5.60&4.57&6.46 \\
TOI 461&9.5&4884&45.6&10/12/19&3.39&4.47&3.64&5.66 \\
TOI 462&11.1&5696&205.4&10/10/19&3.83&5.34&3.97&5.49 \\
TOI 482&14.1&3692&173.8&10/11/19&3.34&4.11&3.59&4.62 \\
TOI 484&12.4&4421&150.0&10/11/19&3.08&4.20&3.72&5.44 \\
TOI 488&12.5&3329&27.4&11/18/19&3.71&4.34&4.18&5.26 \\
TOI 493&12.2&4139&107.4&10/14/19&3.40&4.46&4.28&5.64 \\
TOI 503&9.3&7764&255.4&10/14/19&4.15&4.93&4.33&6.24 \\
TOI 509&8.4&5560&49.0&10/14/19&4.05&4.85&4.61&6.44 \\
TOI 515&14.4&4952&442.8&10/14/19&3.26&4.13&3.10&4.43 \\
TOI 518&10.5&5891&159.8&10/14/19&3.61&3.92&4.04&5.62 \\
TOI 523$^d$ & 9.6 & 4914 & 78.0 & 11/18/19 & 3.47 & 3.94 & 4.17 & 5.59 \\
TOI 524&10.4&6924&293.3&10/10/19&4.43&4.98&4.60&5.84 \\
TOI 526&13.4&3601&70.9&10/10/19&4.16&4.79&3.85&5.63 \\
TOI 530&14.6&3566&148.8&10/13/19&3.38&4.54&3.71&5.20 \\
TOI 532&13.7&3815&135.0&10/10/19&4.07&4.86&4.23&5.93 \\
TOI 538&15.4&3352&133.2&10/11/19&3.58&3.91&2.72&3.53 \\
TOI 544&10.4&4220&41.1&10/10/19&3.70&4.54&4.44&5.93 \\
TOI 554&6.8&6338&45.6&10/11/19&4.23&6.69&3.82&7.78 \\
TOI 556$^e$&11.9&5056&146.9&10/10/19&4.15&4.89&4.14&5.05 \\
TOI 557&12.6&3841&76.0&10/12/19&3.96&4.49&3.69&5.49 \\
TOI 603&10.1&5901&205.9&11/18/19&3.60&5.18&4.14&5.46 \\
TOI 628&10.1&6174&178.7&10/11/19&3.77&4.49&4.09&5.42 \\
TOI 629&8.7&9165&333.4&10/10/19&3.70&5.36&4.65&7.80 \\
TOI 647&10.8&4900&553.6&11/18/19&3.61&4.41&4.23&5.34 \\
TOI 685$^f$  & 10.5 & 5466 & 213.3 &02/05/18& 3.79 & 5.41 & 3.88 & 7.52 \\
TOI 692&9.0&9622&482.0&10/13/19&3.81&5.13&4.12&6.11 \\
TOI 693&11.9&4654&114.8&11/17/19&4.03&4.77&3.73&5.51 \\
TOI 727&12.1&3653&43.0&11/18/19&3.68&4.07&4.14&5.80 \\
TOI 774$^g$  & 11.8 & 6070 & 297.5 &04/18/16& 2.87 & 3.13 & 2.82 & 3.25 \\
TOI 844&12.2&5830&472.3&10/10/19&3.92&4.62&3.76&5.20 \\
TOI 851&11.5&5485&154.5&10/12/19&3.73&4.15&3.49&5.14 \\
TOI 852&11.4&5574&351.4&10/12/19&3.49&4.76&3.70&5.42 \\
TOI 855&11.0&6671&294.4&10/10/19&4.10&4.80&4.00&5.41 \\
TOI 879&9.6&9839&602.7&10/10/19&3.98&4.89&4.19&5.97 \\
TOI 880&9.8&4935&60.7&11/18/19&3.72&4.55&4.48&5.96 \\
TOI 881&10.4&5274&994.7&10/11/19&3.97&4.79&4.20&6.13 \\
TOI 882&10.0&7069&388.0&10/10/19&4.77&5.84&4.59&6.05 \\
TOI 883&9.8&5651&102.6&10/13/19&3.75&4.33&4.25&6.32 \\
TOI 884&10.0&11246&1390.5&10/11/19&3.81&4.86&3.89&5.34 \\
TOI 885&10.9&4692&693.4&11/18/19&3.52&4.15&4.08&6.08 \\
TOI 886&8.3&8844&364.9&10/13/19&3.61&5.07&4.36&6.95 \\
TOI 888&9.8&6822&263.0&10/10/19&4.49&5.66&4.23&5.87 \\
TOI 890&11.3&6935&474.6&11/18/19&3.45&4.12&3.66&4.64 \\
TOI 892&11.3&7723&340.5&11/10/19&3.96&4.67&3.96&5.37 \\
TOI 893&11.6&9856&1241.4&10/13/19&3.71&4.57&3.88&6.64 \\
TOI 894&9.2&9900&661.7&10/13/19&3.79&4.67&4.23&6.53 \\
TOI 895&9.2&5998&96.3&10/11/19&3.38&4.33&3.80&5.59 \\
TOI 896&9.4&6627&156.1&10/10/19&4.68&5.81&4.48&6.34 \\
TOI 897&9.5&6128&165.9&10/11/19&4.18&5.02&4.19&6.47 \\
TOI 898&11.0&5895&479.1&11/18/19&3.65&4.26&3.77&5.26 \\
TOI 938&11.3&5981&215.0&10/12/19&3.48&4.36&3.58&5.32 \\
TOI 939&11.3&6160&352.2&10/12/19&3.49&4.35&3.76&5.45 \\
TOI 941&11.4&5920&263.4&10/10/19&3.66&4.59&4.46&6.29 \\
TOI 943&11.4&6794&397.8&10/10/19&4.20&4.78&4.26&5.48 \\
TOI 944&12.0&7011&938.4&10/10/19&4.13&5.02&4.15&5.38 \\
TOI 950&10.7&6706&211.6&10/10/19&4.30&5.12&4.11&5.82 \\
TOI 952&10.3&7110&459.2&10/12/19&3.64&4.64&3.68&5.36 \\
TOI 957 & 9.0 & 8897 & 280.2 & 10/10/19 & 4.16 & 5.04 & 4.52 & 6.42 \\
TOI 958 & 11.4 & 5745 & 297.1 & 10/11/19 & 2.98 & 3.89 & 3.91 & 4.90 \\
TOI 959&10.7&7491&639.6&10/13/19&3.67&4.70&4.20&6.20 \\
TOI 960&10.7&9385&789.9&11/18/19&4.00&4.48&3.82&5.18 \\
TOI 961&11.1&5924&246.2&10/13/19&3.48&4.40&4.07&5.46 \\
TOI 963&11.1&5815&203.9&10/13/19&3.27&4.35&3.93&5.91 \\
TOI 965&11.0&6110&224.5&11/18/19&3.56&4.05&3.85&5.66 \\
TOI 969&11.3&4249&77.3&10/13/19&2.30&4.20&4.08&6.23 \\
TOI 971&11.0&5743&229.6&11/18/19&3.45&3.95&4.14&6.16 \\
TOI 973&11.9&3435&4153.0&11/18/19&3.73&4.00&3.89&6.20 \\
TOI 977&11.3&6307&6862.4&10/11/19&3.20&4.28&4.20&6.01 \\
TOI 978&10.7&6368&291.6&11/18/19&4.15&5.04&4.25&5.83 \\
TOI 979&10.8&5806&414.9&11/18/19&3.96&4.50&3.71&5.03 \\
TOI 980&10.8&5322&269.6&11/18/19&3.92&4.57&4.27&5.59 \\
TOI 982&10.4&8502&793.4&11/18/19&3.59&4.12&3.95&5.49 \\
TOI 984&10.7&7773&442.8&11/18/19&3.78&4.76&4.08&5.71 \\
TOI 985&10.7&6003&260.6&10/14/19&3.28&4.23&4.22&5.51 \\
TOI 986&10.3&8031&407.5&10/13/19&3.57&4.69&3.83&5.73 \\
TOI 989&10.3&7875&476.8&10/11/19&3.99&5.35&3.90&5.98 \\
TOI 994&10.0&10393&543.1&11/18/19&4.06&4.81&3.78&5.73 \\
TOI 995&10.7&4920&1080.9&10/11/19&3.85&4.99&4.25&6.17 \\
TOI 1002&9.4&8924&943.1&11/18/19&3.17&4.01&4.14&6.36 \\
TOI 1007&9.2&6596&283.3&11/18/19&3.40&4.15&4.16&5.69 \\
TOI 1008&9.3&6699&144.3&10/11/19&3.96&5.37&3.94&6.17 \\
TOI 1012&8.2&8928&296.0&10/14/19&3.87&4.81&4.62&6.61 \\
 TOI 1132 & 9.4 & 7880 & 286.7 & 10/14/19 & 3.90 & 4.70 & 4.44 & 5.99 \\
 TOI 1133 & 9.5 & 6244 & 233.4 & 10/14/19 & 4.86 & 6.52 & 4.68 & 7.41 \\
 TOI 1134 & 9.4 & 6277 & 170.7 & 10/11/19 & 4.18 & 5.02 & 4.24 & 5.20 \\
 TOI 1138 & 9.0 & 9994 & 395.1 & 10/11/19 & 4.53 & 5.25 & 4.67 & 5.53 \\
 TOI 1144$^h$ &  9.2 & 4777 &  37.8 & 06/13/11   & 4.02 & 6.27 & 3.84 & 4.71 \\
 TOI 1145 & 8.4 & 12433 & 438.4 & 10/11/19 & 4.51 & 5.46 & 4.50 & 5.54 \\
 TOI 1149 & 7.8 & 13079 & 632.6 & 10/14/19 & 4.16 & 5.44 & 4.77 & 6.91 \\
 TOI 1152$^i$ & 8.5 & 5485 & 105.7 & 11/16/19 & 3.76 & 4.97 & 4.83 & 6.21 \\
 TOI 1159 & 9.9 & 6592 & 292.8 & 10/10/19 & 4.16 & 4.95 & 4.44 & 5.83 \\
 TOI 1161$^j$ & 10.4 & 7986 & 500 & 06/22/20 & 2.96 & 4.67 & 3.31 & 4.34 \\
 TOI 1162$^k$ & 9.8 & 8730 & 352.0 & 10/15/19 & 3.60 & 4.71 & 4.09 & 5.91 \\
 TOI 1163$^l$ & 9.6 & 9311 & 148.7 & 10/14/19 & 4.64 & 6.35 & 4.36 & 6.67 \\
 TOI 1170&10.5&7734&880.1&10/10/19&3.83&4.41&4.56&5.50 \\
 TOI 1171 & 10.5 & 7550 & 482.8 & 10/12/19 & 4.72 & 6.34 & 4.25 & 5.49 \\
 TOI 1175 & 10.6 & 6229 & 216.5 & 10/11/19 & 3.95 & 4.77 & 4.38 & 5.59 \\
 TOI 1178 & 11.1 & 3897 & 36.7 & 10/10/19 & 3.77 & 4.34 & 4.29 & 5.74 \\
 TOI 1181 & 10.5 & 6122 & 302.8 & 10/14/19 & 3.69 & 4.29 & 3.59 & 5.09 \\
 TOI 1183 & 10.5 & 5599 & 112.9 & 10/11/19 & 3.74 & 4.16 & 3.86 & 5.18 \\
 TOI 1184 & 10.6 & 4534 & 58.6 & 10/10/19 & 3.66 & 4.27 & 4.38 & 5.80 \\
 TOI 1189$^m$ & 10.4 & 5287 & 248.6 & 10/15/19 & 4.11 & 5.12 & 4.31 & 5.44 \\
 TOI 1191 & 10.1 & 6800 & 355.8 & 10/11/19 & 3.87 & 4.75 & 3.78 & 4.90 \\
 TOI 1192 & 10.8 & 6479 & 283 & 10/12/19 & 3.91 & 4.92 & 4.09 & 5.31 \\
 TOI 1195 & 11.0 & 5246 & 500.0 & 10/12/19 & 4.09 & 5.91 & 4.20 & 5.95 \\
 TOI 1196 & 10.8 & 6689 & 424.4 & 10/15/19 & 3.31 & 5.07 & 3.94 & 5.53 \\
 TOI 1197 & 10.8 & 7649 & 405.6 & 10/15/19 & 4.31 & 5.20 & 4.32 & 6.08 \\
 TOI 1201&12.1&3506&37.9&11/10/19&3.81&4.40&3.86&5.91 \\
 TOI 1235 & 10.8 & 3912 & 39.6 & 10/14/19 & 3.22 & 4.07 & 4.04 & 5.61 \\
 TOI 1237$^n$ & 10.6 & 6212 & 243.3 & 06/24/10   & 4.07 & 4.96 & 3.67 & 4.34 \\
 TOI 1241$^o$ & 11.6 & 5826 & 546.5 & 11/18/19 & 3.89 & 4.16 & 3.69 & 4.97 \\
 TOI 1251 & 11.1 & 5273 & 186.0 & 11/09/19 & 3.37 & 3.56 & 3.45 & 4.84 \\
 TOI 1263 & 9.1 & 5098 & 46.6 & 11/16/19 & 3.37 & 4.05 & 4.00 & 5.53 \\
 TOI 1264 & 11.2 & 5040 & 141.8 & 11/16/19 & 3.03 & 3.78 & 3.76 & 4.86 \\
 TOI 1265$^p$ & 10.4 & 6532 & 341.1 & 06/13/11 & 3.96 & 5.71 & 2.76 & 3.15 \\
 TOI 1267$^q$ & 11.9 & 6378 & 980 & 10/25/10   & 2.46 & 3.46 & 2.28 & 3.30 \\
 TOI 1287 & 9.0 & 5891 & 92.7 & 11/18/19 & 3.94 & 4.42 & 4.75 & 6.41 \\
 TOI 1288 & 10.4 & 6180 & 114.9 & 11/17/19 & 3.52 & 4.37 & 4.01 & 5.84 \\
 TOI 1290$^r$ &  9.9 & 5875 & 144.2 & 06/20/10   & 2.62 & 4.36 & 3.32 & 5.60 \\
 TOI 1301 & 11.1 & 4781 & 90.9 & 11/09/19 & 3.44 & 3.73 & 3.86 & 5.45 \\
 TOI 1305$^s$ & 10.6 & 5267 & 664.4 & 11/16/19 & 3.38 & 3.95 & 4.11 & 5.03 \\
 TOI 1306 & 10.5 & 5273 & 364.5 & 11/09/19 & 3.34 & 3.66 & 4.32 & 5.55 \\
 TOI 1307$^t$ & 11.4 & 5010 & 765.3 & 11/18/19 & 3.76 & 4.25 & 3.82 & 5.10 \\
 TOI 1311 & 10.7 & 8153 & 556.8 & 11/18/19 & 3.95 & 4.57 & 4.03 & 5.53 \\
 TOI 1314 & 10.5 & 5155 & 285.4 & 11/17/19 & 3.59 & 4.17 & 4.23 & 5.37 \\
 TOI 1315 & 9.3 & 8321 & 453.7 & 11/17/19 & 4.00 & 4.27 & 4.34 & 6.05 \\
 TOI 1316 & 10.7 & 6556 & 435.0 & 11/18/19 & 3.47 & 4.03 & 4.50 & 5.89 \\
 TOI 1317 & 10.6 & 8542 & 697.1 & 11/17/19 & 3.11 & 4.10 & 4.11 & 5.56 \\
 TOI 1320$^u$ & 10.4 & 6500 & 340.3 & 11/17/19 & 3.71 & 4.08 & 3.94 & 4.85 \\
 TOI 1321 & 10.3 & 9067 & 1139.3 & 11/17/19 & 3.68 & 4.05 & 4.08 & 5.24 \\
 TOI 1323 & 8.3 & 9003 & 301.4 & 11/09/19 & 3.91 & 5.48 & 4.10 & 6.06 \\
 TOI 1324 & 10.5 & 6854 & 339.7 & 11/17/19 & 3.61 & 4.08 & 3.90 & 5.63 \\
 TOI 1327 & 8.5 & 8715 & 457.8 & 11/16/19 & 3.34 & 3.88 & 4.08 & 5.37 \\
 TOI 1328 & 10.7 & 6585 & 397.9 & 11/17/19 & 3.42 & 4.15 & 4.13 & 5.37 \\
 TOI 1329 & 10.5 & 7537 & 398.6 & 11/17/19 & 3.69 & 3.96 & 4.20 & 5.54 \\
 TOI 1334 & 9.6 & 11348 & 1276.3 & 11/18/19 & 3.95 & 4.21 & 4.35 & 5.57 \\
 TOI 1339 & 8.8 & 5461 & 53.6 & 11/18/19 & 3.51 & 4.20 & 4.63 & 6.33 \\
 TOI 1342$^v$ & 10.8 & 5869 & -- & 11/18/19 & 3.83 & 4.24 & 3.34 & 4.78 \\
 TOI 1353 & 10.3 & 9706 & 464.2 & 11/17/19 & 3.60 & 4.42 & 4.10 & 5.62 \\
 TOI 1354 & 8.8 & 9224 & 245.8 & 11/09/19 & 4.10 & 5.48 & 3.62 & 6.75 \\
 TOI 1355 & 8.7 & 9218 & 248.1 & 11/09/19 & 3.47 & 4.13 & 4.12 & 5.92 \\
 TOI 1356 & 9.0 & 12308 & 4772.6 & 11/09/19 & 3.89 & 4.75 & 4.34 & 6.03 \\
 TOI 1357 & 10.6 & 7387 & 254.4 & 11/17/19 & 3.68 & 4.12 & 3.87 & 5.79 \\
 TOI 1358 & 8.6 & 10445 & 356.7 & 11/09/19 & 3.68 & 4.37 & 4.19 & 5.43 \\
 TOI 1360 & 10.1 & 8999 & 533.6 & 11/18/19 & 4.00 & 4.25 & 4.40 & 5.35 \\
 TOI 1362 & 10.2 & 8129 & 975.5 & 11/17/19 & 3.55 & 4.15 & 4.22 & 5.72 \\
 TOI 1364 & 9.2 & 9598 & 595.3 & 11/18/19 & 4.11 & 4.59 & 4.53 & 5.76 \\
 TOI 1366 & 8.9 & 5717 & 118.6 & 11/10/19 & 3.45 & 4.30 & 3.93 & 6.13 \\
 TOI 1367 & 9.6 & 6550 & 195.9 & 11/10/19 & 3.62 & 4.41 & 4.01 & 5.79 \\
 TOI 1368 & 9.6 & 5717 & 158.2 & 11/16/19 & 3.25 & 3.90 & 3.70 & 5.04 \\
 TOI 1369 & 10.5 & 9139 & 516.3 & 11/09/19 & 3.67 & 3.96 & 4.23 & 4.95 \\
 TOI 1370 & 9.4 & 8728 & 275.5 & 11/09/19 & 3.52 & 4.13 & 4.15 & 5.70 \\
 TOI 1376 & 10.7 & 6026 & 240.0 & 11/17/19 & 3.59 & 4.03 & 4.09 & 5.88 \\
 TOI 1378 & 9.8 & 8224 & 547.4 & 11/09/19 & 3.83 & 4.18 & 4.26 & 5.55 \\
 TOI 1381 & 10.5 & 9976 & 737.1 & 11/17/19 & 3.80 & 4.10 & 4.19 & 5.20 \\
 TOI 1384 & 10.5 & 5773 & 235.2 & 11/17/19 & 3.82 & 4.15 & 4.20 & 5.53 \\
 TOI 1385$^w$ & -- & -- & 324.3 & 11/09/19 & 3.82 & 4.23 & 4.11 & 5.87 \\
 TOI 1386 & 10.5 & 5769 & 146.9 & 11/17/19 & 3.64 & 4.30 & 3.99 & 5.44 \\
 TOI 1387 & 9.1 & 7976 & 239.7 & 11/09/19 & 3.71 & 5.26 & 3.97 & 6.11 \\
 TOI 1391 & 11.0 & 5256 & 115.7 & 11/17/19 & 3.75 & 4.30 & 4.31 & 5.71 \\
 TOI 1393$^x$ & 10.4 & 7000 & 278.8 & 11/17/19 & 3.67 & 4.19 & 4.44 & 5.73 \\
 TOI 1394 & 9.8 & 6294 & 195.0 & 11/17/19 & 3.67 & 4.25 & 4.44 & 6.34 \\
 TOI 1397 & 10.6 & 6357 & 195.0 & 11/17/19 & 3.32 & 3.98 & 4.08 & 6.08 \\
 TOI 1398 & 8.6 & 9623 & 648.4 & 11/09/19 & 4.11 & 5.25 & 4.42 & 6.28 \\
 TOI 1399 & 10.6 & 6483 & 252.4 & 11/09/19 & 3.47 & 3.86 & 4.21 & 5.22 \\
 TOI 1400 & 11.3 & 6004 & 370.8 & 11/18/19 & 3.40 & 3.68 & 4.06 & 5.05 \\
 TOI 1401 & 11.4 & 6403 & 383.2 & 11/18/19 & 3.36 & 4.25 & 3.83 & 4.32 \\
 TOI 1402 & 10.8 & 7111 & 576.7 & 11/18/19 & 3.31 & 3.74 & 4.10 & 4.67 \\
 TOI 1405$^y$ & 9.1 & 5195 & -- & 11/09/19 & 3.70 & 4.05 & 4.30 & 5.36 \\
 TOI 1407$^z$ &  8.1 & 6129 &  80.9 & 09/28/15   & 3.06 & 3.97 & 2.97 & 3.92 \\
 TOI 1554$^{aa}$ & 11.5 & 5497 &   194 & 09/17/10   & 2.95 & 4.36 & 3.48 & 4.85 \\
 TOI 1905$^{ab}$ & 11.2 & 4251 & 64.7 & 04/11/17 & 3.67 & 3.97 & 4.19 & 6.05\\
 TIC125192758 & 14.7 & 5384 & 1100.5 & 11/17/19 & 3.71 & 4.11 & 3.78 & 4.31 \\
 %
\enddata
\tablenotetext{a}{HATS-3}
\tablenotetext{b}{HATS-14}
\tablenotetext{c}{Match in Gaia DR2, but no parallax. Distance from ExoFOP}
\tablenotetext{d}{Duplicate entry in TIC; TIC 93125144=TIC 708525747; TOI 523 assigned to TIC 93125144 but Gaia parameters are TIC 708525747}
\tablenotetext{e}{K2-78 - possible false positive}
\tablenotetext{f}{K2-261}
\tablenotetext{g}{WASP-55}
\tablenotetext{h}{HAT-P-11/Kepler-3: data taken with DSSI in 692nm and 880nm filters}
\tablenotetext{i}{Duplicate entry in TIC: TOI 1152 assigned to TIC 237184773; Gaia DR2 has two sources 1" apart Distance from Gaia DR2 query: Gaia DR2 2094001134684220800 and 2094001138979921408.  Distance is from Gaia DR2 2094001134684220800}
\tablenotetext{j}{Kepler-13; TOI 1161 associated with TIC 158324245 but this is resolved by Gaia as two stars: Gaia DR2 2130632159134827392 and 2130632159130638464 which are associated with TIC 1717079071 and TIC 1717079066. Data taken in 562nm and 880nm}
\tablenotetext{k}{Duplicate entry in TIC; TIC 13419950=TIC 1969293164; TOI 1162 assigned to TIC 13419950 but Gaia parameters are TIC 1969293164}
\tablenotetext{l}{Duplicate entry in TIC; TIC 375542276=TIC 1847139036; TOI 1163 assigned to TIC 1847139036 but Gaia parameters are TIC 1847139036}
\tablenotetext{m}{No Gaia information in TIC, Distance from Gaia DR2 query: Gaia DR2 2019824786095520128}
\tablenotetext{n}{Kepler-25: data taken with DSSI in 692nm and 880nm filters}
\tablenotetext{o}{KOI5}
\tablenotetext{p}{HAT-P-7/Kepler-2: data taken with DSSI in 692nm and 880nm filters}
\tablenotetext{q}{Kepler-14: data taken with DSSI in 692nm and 880nm filters; Gaia magnitude calculated from B-V; distance from \citet{2011ESS.....2.1917B}}
\tablenotetext{r}{Kepler-68: data taken with DSSI in 562nm and 692nm filters}
\tablenotetext{s}{TOI 1305(TIC 232679662) = TOI 1172 (TIC 1717732429): possible Nearby EB confusing the signal}
\tablenotetext{t}{No Gaia information in TIC, Distance from Gaia DR2 query: Gaia DR2 2155491910878597376}
\tablenotetext{u}{TESS magnitude given instead of Gaia magnitude; effective temperature and distance from ExoFOP}
\tablenotetext{v}{TIC/Gaia DR2 has Gaia magnitude but no parallax}
\tablenotetext{w}{No Gaia DR2 values in TIC; TOI 1385 is HD~211030 and a is known double star. }
\tablenotetext{x}{TOI 1393 associated with TIC 430528566 but this is resolved by Gaia as two stars: Gaia DR2 2004338577092552192 and 2004338572785772800 which are associated with TIC 2014876481 and TIC 201487661}
\tablenotetext{y}{Gaia DR2 only has magnitude with no parallax measurement}
\tablenotetext{z}{K2-167}
\tablenotetext{aa}{Kepler-63: data taken with DSSI in 562nm and 692nm filters}
\tablenotetext{ab}{WASP-107/K2-235; Data taken in 562nm and 832nm}
\end{deluxetable}

\startlongtable
\begin{deluxetable}{lccccccc}
\tablecaption{TESS Stars with Close Companions}
\tablehead{
\colhead{} & \multicolumn{3}{c}{562 nm} &
\multicolumn{3}{c}{832 nm} & \colhead{} \\
\colhead{Target} & 
\colhead{Sep (")} & 
\colhead{PA (deg)} & 
\colhead{$\Delta$Mag} & 
\colhead{Sep (")} & 
\colhead{PA (deg)} & 
\colhead{$\Delta$Mag} &
\colhead{Sep (au)}
}
\startdata
TOI 123&1.296&295.06&2.24&1.285&294.96&1.96 & 212\\
TOI 172&1.116&320.59&4.81&1.116&320.59&4.81 & 383\\
TOI 309&--&--&--&0.326&74.7&2.31&112.9\\
TOI 462&0.167&197.15&0.34&0.169&196.37&0.54 & 34 \\
TOI 482&--&--&--&0.398&267.34&2.43 & 69\\
TOI 851&--&--&--&1.839&255.39&5.53 & 646 \\
TOI 890&0.424&146.69&0.71&0.423&146.65&0.67 & 201 \\
TOI 894&0.592&331.11&3.13&0.596&330.87&2.41 & 394 \\
TOI 898&1.291&37.49&2.46&1.289&36.23&2.28 & 618\\
TOI 944&2.025&112.78&2.19&2.006&112.48&1.96 & 1882 \\
TOI 952&--&--&--&1.162&135.34&4.74 & 534\\
TOI 979&0.084&241.02&0.57&0.084&239.97&0.55 & 35 \\
TOI 984&0.281&187.29&2.30&0.283&186.51&2.15 & 125 \\
TOI 994&--&--&--&1.371&301.43&6.59 & 745 \\
TOI 1008&0.487&129.85&0.33&0.484&129.70&0.33 & 70 \\
TOI 1133 &  0.548 &  103.305 &   3.05 & 0.542 &  102.810 &   2.75 &  127 \\
TOI 1152 &  1.587 &  165.670 &   1.04 &  1.593 &  165.311 &   0.73 &  168 \\
TOI 1161 & 1.144 & 279.7 & 0.195 & 1.144 & 279.7 & 0.140 & 572\\
TOI 1162 &  1.466 &  132.344 &   2.11 &  1.463 &  132.078 &   1.40 &  516 \\
TOI 1163$^{c}$ &  1.671 &  114.622 &   4.93 &  1.662 &  114.146 &   3.83 &  248 \\
 TOI 1163$^{c}$ &  0.614 &  216.892 &   3.02 &  0.611 &  215.724 &   2.28 &   91 \\
 TOI 1183 &  0.933 &  225.474 &   0.66 &  0.929 &  224.297 &   0.65 &  105 \\
 TOI 1189 &  0.975 &  319.317 &   1.40 &  0.975 &  319.123 &   1.54 &  242 \\
 TOI 1191 &     -- &       -- &     -- &  0.046 &  323.172 &   1.54 &   16 \\
 TOI 1192 &  0.181 &  200.617 &   3.22 &  0.194 &  201.073 &   3.00 &   53 \\
 TOI 1196 &  1.674 &  199.730 &   2.91 &  1.680 &  198.751 &   2.27 &  712 \\
 TOI 1197 &  1.609 &   26.153 &   4.68 &  1.597 &   25.054 &   3.83 &  650 \\
TOI 1241 &     -- &       -- &     -- &  0.067 &  320.252 &   1.92 &   37 \\
 TOI 1251 &  0.219 &  251.551 &   1.66 &  0.221 &  250.112 &   1.36 &   41 \\
 TOI 1264 &  1.903 &  313.670 &   2.65 &  1.896 &  313.544 &   1.82 &  269 \\
 TOI 1267 &  0.287 &    144.0 &    0.69 &  0.287 &  144.0 & 0.594 & 281\\ 
 TOI 1287 &     -- &       -- &     -- &  0.104 &  346.121 &   3.16 &   10 \\
 TOI 1288$^b$ &     -- &       -- &     -- &  1.123 &  289.478 &   5.90 & 129 \\
 TOI 1288$^b$ &     -- &       -- &     -- &  0.065 &  312.561 &   2.57 &   7 \\
 TOI 1305 &  1.080 &  234.860 &   0.61 &  1.073 &  233.790 &   0.42 &  715 \\
 TOI 1307 &  0.337 &   32.769 &   3.12 &  0.330 &   34.734 &   3.05 &  255 \\
 TOI 1320 &  0.321 &   49.544 &   0.38 &  0.319 &   48.485 &   0.29 &  109 \\
 TOI 1324$^{c}$ &  1.488 &  194.209 &   3.52 &  1.480 &  187.164 &   3.11 &  504 \\
 TOI 1342$^b$ &  0.377 &  163.202 &   0.76 &  0.380 &  162.798 &   0.68 &      355$^a$ \\
TOI 1356 &     -- &       -- &     -- &  0.263 &  288.715 &   4.10 & 1255 \\
TOI 1364 &  2.375 &  110.736 &   5.26 &  2.363 &  110.751 &   3.85 & 1410 \\
TOI 1385 &  0.252 &  187.738 &   0.38 &  0.254 &  187.044 &   0.32 &   82 \\
TOI 1387$^{c}$ &  2.274 &  344.009 &   3.64 &  2.319 &  345.589 &   2.71 &  550 \\
TOI 1401 &  0.200 &   45.363 &   0.99 &  0.197 &   43.253 &   0.96 &   76 \\
TOI 1405 &  0.386 &   42.201 &   1.42 &  0.385 &   40.538 &   1.62 &       377$^a$ \\
\enddata
\tablenotetext{a}{This value is calculated using spectroscopic parallax estimated from the Table 1 stellar parameters.}
\tablenotetext{b}{The filters used were 'r' and 'i', instead of 562 nm and 832 nm, respectively. Two companions were detected}
\tablenotetext{c}{These TOIs were observed multiple times in which the companion was detected. Variations between filters \& observations were: 
TOI 1163 (9 times) Sep$\pm$0.042;PA $\pm$0.041; $\Delta$Mag$\pm$0.52 (the ranges given include both companion stars).
TOI 1387(2 times) Sep$\pm$0.021;PA $\pm$0.053; $\Delta$Mag$\pm$0.45.
TOI 1324(4 times) Sep$\pm$0.009;PA $\pm$3.29; $\Delta$Mag$\pm$0.13.}

\end{deluxetable}
\startlongtable
 \centerwidetable
\begin{deluxetable*}{ccccccc|ccc|ccc|c}
\tablecaption{Stellar parameters and companion space observable with speckle imaging \label{tab:comptable}}
\tabletypesize{\footnotesize}
\tablecolumns{14}
\tablewidth{0pt}
\tablehead{
\colhead{TOI} & 
\colhead{$T_{\mathrm{eff}}$} &
\colhead{$V$} & 
\colhead{$TESS$} & 
\colhead{D} & 
\colhead{SpT$^{a}$} &
\colhead{$M_{\mathrm{interp}}$$^{b}$} & 
\multicolumn{2}{c}{Companion SpT at:} &
\colhead{Comp.} &
\colhead{Min Sep} & 
\colhead{Max Sep} &
\colhead{Distr.} &
\colhead{Speckle} 
\vspace{-2pt} \\
\cline{8-9}
 \colhead{} &
 \colhead{$(K)$} &  
 \colhead{(mag)} & 
 \colhead{(mag)} & 
 \colhead{(pc)} & 
 \colhead{} &
 \colhead{($M_{\sun}$)} &
 \colhead{$\Delta m_v = 4$} & 
 \colhead{$\Delta m_v = 6$} & 
 \colhead{Frac.} &
 \colhead{(AU)} & 
 \colhead{(AU)} & 
 \colhead{Frac.} & 
 \colhead{Frac.}
 }
\startdata
TOI 103 & 6371 & 12.43 & 11.52 & 411.21 & F6V & 1.241 & K5V & M1V & 0.569 & 16.62 & 493.45 & 0.362 & 0.206 \\
TOI 109 & 5361 & 14.05 & 13.24 & 513.01 & G9V & 0.928 & M0.5V & M3V & 0.522 & 20.73 & 615.61 & 0.358 & 0.187 \\
TOI 123 & 6356 & 8.32 & 7.88 & 161.5 & F6V & 1.257 & K5V & M1V & 0.569 & 6.53 & 193.8 & 0.368 & 0.21 \\
TOI 172 & 5759 & 11.3 & 10.74 & 342.83 & G2V & 1.015 & K9V & M2.5V & 0.514 & 13.85 & 411.4 & 0.365 & 0.188 \\
TOI 254 & 6101 & 10.26 & 9.83 & 133.31 & F9V & 1.101 & K6V & M1.5V & 0.557 & 5.39 & 159.97 & 0.366 & 0.204 \\
TOI 260 & 4049 & 9.9 & 8.5 & 20.19 & K7V & 0.645 & M3.5V & M4.5V & 0.589 & 0.82 & 24.22 & 0.376 & 0.222 \\
TOI 266 & 5784 & 10.07 & 9.46 & 101.69 & G2V & 0.997 & K9V & M2.5V & 0.514 & 4.11 & 122.03 & 0.362 & 0.186 \\
TOI 278 & 2950 & 16.45 & 13.17 & 44.4 & M5.5V & 0.096 & M9V & M9V & 1.0 & 1.79 & 53.28 & 0.884 & 0.884 \\
TOI 309 & 5407 & 13.06 & 12.33 & 345.29 & G9V & 0.923 & M0.5V & M3V & 0.522 & 13.95 & 414.35 & 0.365 & 0.191 \\
TOI 316 & 4245 & 14.38 & 13.12 & 275.02 & K6V & 0.684 & M3V & M4.5V & 0.571 & 11.11 & 330.02 & 0.395 & 0.226 \\
TOI 329 & 5560 & 11.26 & 10.69 & 284.39 & G6V & 0.968 & M0V & M2.5V & 0.515 & 11.49 & 341.27 & 0.367 & 0.189 \\
TOI 390 & 6321 & 10.33 & 9.89 & 167.31 & F6V & 1.175 & K5V & M1V & 0.569 & 6.76 & 200.77 & 0.368 & 0.21 \\
TOI 438 & 5210 & 10.24 & 9.45 & 72.46 & K1V & 0.85 & M1V & M3V & 0.53 & 2.93 & 86.96 & 0.355 & 0.188 \\
TOI 461 & 4884 & 9.78 & 8.87 & 45.56 & K3V & 0.791 & M2V & M3.5V & 0.548 & 1.84 & 54.67 & 0.41 & 0.225 \\
TOI 462 & 5695 & 11.66 & 10.66 & 205.37 & G4V & 0.995 & M0V & M2.5V & 0.511 & 8.3 & 246.44 & 0.369 & 0.189 \\
TOI 482 & 3692 & 14.94 & 13.08 & 173.82 & M1V & 0.525 & M4.5V & M5.5V & 0.677 & 7.02 & 208.58 & 0.411 & 0.278 \\
TOI 484 & 4421 & 12.64 & 11.69 & 149.95 & K5V & 0.711 & M3V & M4V & 0.568 & 6.06 & 179.94 & 0.414 & 0.235 \\
TOI 488 & 3329 & 13.74 & 11.2 & 27.36 & M3.5V & 0.359 & M5.5V & M7V & 0.866 & 1.11 & 32.84 & 0.604 & 0.523 \\
TOI 493 & 4139 & 12.55 & 11.45 & 107.36 & K7V & 0.654 & M3.5V & M4.5V & 0.589 & 4.34 & 128.83 & 0.419 & 0.247 \\
TOI 503 & 7764 & 9.4 & 9.2 & 255.42 & A7V & 1.695 & K1V & K6V & 0.551 & 10.32 & 306.5 & 0.266 & 0.147 \\
TOI 509 & 5560 & 8.58 & 7.93 & 48.97 & G6V & 0.968 & M0V & M2.5V & 0.515 & 1.98 & 58.77 & 0.342 & 0.176 \\
TOI 515 & 4952 & 14.49 & 13.8 & 442.8 & K2V & 0.803 & M1.5V & M3.5V & 0.539 & 17.89 & 531.36 & 0.361 & 0.194 \\
TOI 518 & 5891 & 10.9 & 10.14 & 159.81 & G1V & 1.048 & K8V & M2V & 0.52 & 6.46 & 191.78 & 0.368 & 0.191 \\
TOI 523 & 4692 & 10.71 & 9.58 & 70.65 & K4V & 0.744 & M2.5V & M4V & 0.566 & 2.85 & 84.78 & 0.418 & 0.236 \\
TOI 524 & 6923 & 10.47 & 10.09 & 293.27 & F1V & 1.459 & K3V & K9V & 0.561 & 11.85 & 351.92 & 0.269 & 0.151 \\
TOI 526 & 3601 & 14.31 & 12.31 & 70.93 & M1.5V & 0.506 & M4.5V & M5.5V & 0.696 & 2.87 & 85.12 & 0.539 & 0.375 \\
TOI 530 & 3566 & 15.4 & 13.53 & 148.76 & M2V & 0.487 & M4.5V & M5.5V & 0.703 & 6.01 & 178.51 & 0.436 & 0.306 \\
TOI 532 & 3815 & 14.41 & 12.68 & 135.05 & M0.5V & 0.568 & M4V & M5V & 0.653 & 5.46 & 162.06 & 0.451 & 0.295 \\
TOI 538 & 3352 & 16.54 & 14.15 & 133.21 & M3V & 0.363 & M5V & M6.5V & 0.762 & 5.38 & 159.85 & 0.453 & 0.345 \\
TOI 544 & 4220 & 10.78 & 9.65 & 41.12 & K6V & 0.677 & M3V & M4.5V & 0.571 & 1.66 & 49.34 & 0.407 & 0.232 \\
TOI 554 & 6337 & 6.91 & 6.44 & 45.62 & F6V & 1.3 & K5V & M1V & 0.569 & 1.84 & 54.74 & 0.339 & 0.193 \\
TOI 556 & 5055 & 12.18 & 11.31 & 146.9 & K2V & 0.803 & M1.5V & M3.5V & 0.539 & 5.94 & 176.29 & 0.367 & 0.198 \\
TOI 557 & 3841 & 13.34 & 11.64 & 75.96 & M0V & 0.542 & M4V & M5V & 0.614 & 3.07 & 91.15 & 0.531 & 0.326 \\
TOI 603 & 5900 & 10.31 & 9.74 & 205.92 & G0V & 1.075 & K7V & M2V & 0.524 & 8.32 & 247.11 & 0.369 & 0.193 \\
TOI 628 & 6174 & 10.11 & 9.66 & 178.68 & F8V & 1.264 & K6V & M1.5V & 0.57 & 7.22 & 214.42 & 0.369 & 0.21 \\
TOI 629 & 9165 & 8.74 & 8.69 & 333.38 & A1V & 2.138 & G8V & K4V & 0.533 & 13.47 & 400.05 & 0.272 & 0.145 \\
TOI 647 & 4900 & 11.15 & 10.23 & 553.63 & K3V & 0.792 & M2V & M3.5V & 0.548 & 22.37 & 664.35 & 0.359 & 0.196 \\
TOI 685 & 5466 & 10.61 & 9.96 & 213.28 & G8V & 0.931 & M0.5V & M3V & 0.515 & 8.62 & 255.94 & 0.369 & 0.19 \\
TOI 692 & 9622 & 9.02 & 9.02 & 481.97 & A0V & 2.284 & G5V & K4V & 0.531 & 19.47 & 578.36 & 0.278 & 0.148 \\
TOI 693 & 4654 & 12.0 & 11.22 & 114.76 & K4V & 0.749 & M2.5V & M4V & 0.566 & 4.64 & 137.72 & 0.418 & 0.236 \\
TOI 727 & 3653 & 12.68 & 11.0 & 42.97 & M1.5V & 0.495 & M4.5V & M5.5V & 0.696 & 1.74 & 51.57 & 0.585 & 0.407 \\
TOI 774 & 6070 & 11.75 & 11.35 & 297.46 & F9V & 1.114 & K6V & M1.5V & 0.557 & 12.02 & 356.95 & 0.367 & 0.204 \\
TOI 844 & 5829 & 12.12 & 11.8 & 472.3 & G1V & 1.05 & K8V & M2V & 0.52 & 19.08 & 566.76 & 0.359 & 0.187 \\
TOI 851 & 5485 & 11.31 & 11.0 & 154.48 & G8V & 0.814 & M0.5V & M3V & 0.515 & 6.24 & 185.37 & 0.368 & 0.19 \\
TOI 852 & 5574 & 11.64 & 10.95 & 351.43 & G6V & 0.969 & M0V & M2.5V & 0.515 & 14.2 & 421.71 & 0.365 & 0.188 \\
TOI 855 & 6671 & 11.18 & 10.65 & 294.42 & F4V & 1.288 & K4V & M0.5V & 0.57 & 11.9 & 353.3 & 0.367 & 0.209 \\
TOI 879 & 9839 & 9.55 & 9.53 & 602.73 & A0V & 2.318 & G5V & K4V & 0.531 & 24.35 & 723.27 & 0.281 & 0.149 \\
TOI 880 & 4935 & 10.1 & 9.26 & 60.67 & K3V & 0.813 & M2V & M3.5V & 0.548 & 2.45 & 72.8 & 0.349 & 0.191 \\
TOI 881 & 5274 & 10.64 & 9.79 & 994.73 & K0V & 0.868 & M1V & M3V & 0.528 & 40.19 & 1193.68 & 0.336 & 0.177 \\
TOI 882 & 7069 & 10.07 & 9.74 & 387.97 & F1V & 1.517 & K3V & K9V & 0.561 & 15.68 & 465.56 & 0.275 & 0.154 \\
TOI 883 & 5651 & 9.96 & 9.37 & 102.64 & G5V & 0.998 & M0V & M2.5V & 0.509 & 4.15 & 123.16 & 0.363 & 0.185 \\
TOI 884 & 11246 & 9.91 & 9.95 & 1390.0 & B9V & 3.171 & G1V & K3V & 0.519 & 56.17 & 1668.0 & 0.287 & 0.149 \\
TOI 885 & 4692 & 11.46 & 10.18 & 693.39 & K4V & 0.744 & M2.5V & M4V & 0.566 & 28.02 & 832.07 & 0.344 & 0.195 \\
TOI 886 & 8844 & 8.34 & 8.27 & 364.93 & A2V & 2.051 & G8V & K5V & 0.539 & 14.75 & 437.91 & 0.274 & 0.148 \\
TOI 888 & 6822 & 10.02 & 9.53 & 263.02 & F2V & 1.441 & K3V & M0V & 0.563 & 10.63 & 315.63 & 0.267 & 0.15 \\
TOI 890 & 6935 & 11.14 & 10.94 & 474.56 & F1V & 1.481 & K3V & K9V & 0.561 & 19.18 & 569.48 & 0.278 & 0.156 \\
TOI 892 & 7723 & 11.43 & 10.97 & 340.54 & A7V & 1.566 & K1V & K6V & 0.551 & 13.76 & 408.65 & 0.272 & 0.15 \\
TOI 893 & 9856 & 11.92 & 11.48 & 1241.42 & A0V & 2.32 & G5V & K4V & 0.531 & 50.16 & 1489.7 & 0.287 & 0.152 \\
TOI 894 & 9900 & 9.11 & 9.17 & 661.74 & A0V & 2.325 & G5V & K4V & 0.531 & 26.74 & 794.09 & 0.282 & 0.15 \\
TOI 895 & 5998 & 9.35 & 8.8 & 96.3 & F9V & 1.104 & K6V & M1.5V & 0.557 & 3.89 & 115.57 & 0.361 & 0.201 \\
TOI 896 & 6627 & 9.48 & 9.09 & 156.14 & F4V & 1.331 & K4V & M0.5V & 0.57 & 6.31 & 187.37 & 0.368 & 0.21 \\
TOI 897 & 6128 & 9.74 & 9.11 & 165.94 & F8V & 1.167 & K6V & M1.5V & 0.57 & 6.71 & 199.13 & 0.368 & 0.21 \\
TOI 898 & 5895 & 10.83 & 10.57 & 479.08 & G1V & 1.074 & K8V & M2V & 0.52 & 19.36 & 574.9 & 0.359 & 0.187 \\
TOI 938 & 5981 & 11.39 & 10.85 & 215.03 & F9V & 1.063 & K6V & M1.5V & 0.557 & 8.69 & 258.04 & 0.369 & 0.206 \\
TOI 939 & 6159 & 11.35 & 10.93 & 352.2 & F8V & 1.176 & K6V & M1.5V & 0.57 & 14.23 & 422.63 & 0.365 & 0.208 \\
TOI 941 & 5920 & 11.46 & 11.0 & 263.39 & G0V & 1.098 & K7V & M2V & 0.524 & 10.64 & 316.06 & 0.368 & 0.193 \\
TOI 943 & 6794 & 11.2 & 11.04 & 397.79 & F2V & 1.392 & K3V & M0V & 0.563 & 16.07 & 477.35 & 0.363 & 0.204 \\
TOI 944 & 7011 & 11.99 & 11.38 & 938.4 & F1V & 1.492 & K3V & K9V & 0.561 & 37.92 & 1126.08 & 0.285 & 0.16 \\
TOI 950 & 6706 & 11.12 & 10.21 & 211.59 & F3V & 1.384 & K4V & M0V & 0.564 & 8.55 & 253.91 & 0.369 & 0.208 \\
TOI 952 & 7110 & 10.37 & 10.07 & 459.16 & F1V & 1.536 & K3V & K9V & 0.561 & 18.55 & 550.99 & 0.277 & 0.156 \\
TOI 957 & 8897 & 9.04 & 8.96 & 280.19 & A2V & 2.058 & G8V & K5V & 0.539 & 11.32 & 336.23 & 0.268 & 0.145 \\
TOI 958 & 5745 & 11.43 & 10.96 & 297.12 & G2V & 1.008 & K9V & M2.5V & 0.514 & 12.01 & 356.54 & 0.367 & 0.189 \\
TOI 959 & 7491 & 10.85 & 10.65 & 639.56 & A8V & 1.67 & K2V & K7V & 0.555 & 25.84 & 767.48 & 0.282 & 0.157 \\
TOI 960 & 9385 & 10.58 & 10.67 & 789.89 & A1V & 2.257 & G8V & K4V & 0.533 & 31.92 & 947.87 & 0.284 & 0.152 \\
TOI 961 & 5924 & 11.46 & 10.66 & 246.18 & G0V & 1.138 & K7V & M2V & 0.524 & 9.95 & 295.41 & 0.368 & 0.193 \\
TOI 963 & 5814 & 11.0 & 10.64 & 203.87 & G2V & 1.039 & K9V & M2.5V & 0.514 & 8.24 & 244.65 & 0.369 & 0.19 \\
TOI 965 & 6110 & 11.1 & 10.61 & 224.49 & F8V & 1.141 & K6V & M1.5V & 0.57 & 9.07 & 269.39 & 0.369 & 0.21 \\
TOI 969 & 4249 & 11.65 & 10.54 & 77.26 & K6V & 0.685 & M3V & M4.5V & 0.571 & 3.12 & 92.71 & 0.419 & 0.239 \\
TOI 971 & 5743 & 11.17 & 10.52 & 229.6 & G3V & 1.008 & K9V & M2.5V & 0.512 & 9.28 & 275.51 & 0.369 & 0.189 \\
TOI 973 & 3435 & 13.69 & 10.58 & 4153.0 & M3V & 0.401 & M5V & M6.5V & 0.762 & 167.81 & 4983.6 & 0.041 & 0.032 \\
TOI 977 & 6307 & 11.62 & 10.52 & 6862.39 & F6V & 1.237 & K5V & M1V & 0.569 & 277.29 & 8234.87 & 0.233 & 0.133 \\
TOI 978 & 6368 & 10.58 & 10.34 & 291.63 & F6V & 1.262 & K5V & M1V & 0.569 & 11.78 & 349.96 & 0.367 & 0.209 \\
TOI 979 & 5806 & 10.72 & 10.35 & 414.88 & G2V & 1.038 & K9V & M2.5V & 0.514 & 16.76 & 497.85 & 0.362 & 0.186 \\
TOI 980 & 5322 & 10.93 & 10.31 & 269.61 & G9V & 0.891 & M0.5V & M3V & 0.522 & 10.89 & 323.53 & 0.368 & 0.192 \\
TOI 982 & 8502 & 10.58 & 10.31 & 793.41 & A3V & 1.985 & G8V & K5V & 0.541 & 32.06 & 952.09 & 0.284 & 0.154 \\
TOI 984 & 7773 & 10.56 & 10.28 & 442.79 & A7V & 1.749 & K1V & K6V & 0.551 & 17.89 & 531.35 & 0.277 & 0.152 \\
TOI 985 & 6002 & 10.8 & 10.25 & 260.64 & F9V & 1.12 & K6V & M1.5V & 0.557 & 10.53 & 312.77 & 0.368 & 0.205 \\
TOI 986 & 8031 & 10.26 & 10.18 & 407.54 & A6V & 1.709 & K0V & K6V & 0.547 & 16.47 & 489.05 & 0.276 & 0.151 \\
TOI 989 & 7875 & 10.24 & 10.11 & 476.81 & A7V & 1.789 & K1V & K6V & 0.551 & 19.27 & 572.17 & 0.278 & 0.153 \\
TOI 994 & 10393 & 10.09 & 10.03 & 543.15 & B9.5V & 2.399 & G2V & K3V & 0.523 & 21.95 & 651.78 & 0.28 & 0.146 \\
TOI 995 & 4920 & 11.25 & 10.08 & 1080.88 & K3V & 0.796 & M2V & M3.5V & 0.548 & 43.67 & 1297.06 & 0.313 & 0.171 \\
TOI 1002 & 8924 & 9.44 & 9.3 & 943.11 & A2V & 2.067 & G8V & K5V & 0.539 & 38.11 & 1131.73 & 0.286 & 0.154 \\
TOI 1007 & 6596 & 9.37 & 8.88 & 283.29 & F4V & 1.368 & K4V & M0.5V & 0.57 & 11.45 & 339.95 & 0.367 & 0.209 \\
TOI 1008 & 6699 & 9.26 & 8.94 & 144.3 & F3V & 1.415 & K4V & M0V & 0.564 & 5.83 & 173.16 & 0.252 & 0.142 \\
TOI 1012 & 8928 & 8.18 & 8.13 & 295.96 & A2V & 2.068 & G8V & K5V & 0.539 & 11.96 & 355.15 & 0.27 & 0.145 \\
TOI 1132 & 7880 & 9.38 & 9.23 & 286.67 & A7V & 1.779 & K1V & K6V & 0.551 & 11.58 & 344.0 & 0.269 & 0.148 \\
TOI 1133 & 6244 & 9.56 & 9.11 & 233.37 & F7V & 1.212 & K6V & M1V & 0.572 & 9.43 & 280.04 & 0.369 & 0.211 \\
TOI 1134 & 6277 & 9.55 & 9.04 & 170.73 & F7V & 1.225 & K6V & M1V & 0.572 & 6.9 & 204.88 & 0.368 & 0.211 \\
TOI 1138 & 9994 & 8.98 & 8.91 & 395.12 & A0V & 2.336 & G5V & K4V & 0.531 & 15.97 & 474.14 & 0.275 & 0.146 \\
TOI 1144 & 4777 & 9.46 & 8.51 & 37.76 & K3V & 0.775 & M2V & M3.5V & 0.548 & 1.53 & 45.32 & 0.404 & 0.221 \\
TOI 1145 & 12433 & 7.55 & 8.5 & 438.4 & B8V & 3.386 & F6V & K0V & 0.524 & 17.71 & 526.08 & 0.277 & 0.145 \\
TOI 1149 & 13079 & 7.87 & 7.92 & 632.55 & B8V & 3.586 & F6V & K0V & 0.524 & 25.56 & 759.06 & 0.282 & 0.148 \\
TOI 1152 & 5485 & 7.99 & 7.4 & 105.7 & G8V & 0.938 & M0.5V & M3V & 0.515 & 4.27 & 126.84 & 0.363 & 0.187 \\
TOI 1159 & 6592 & 10.07 & 9.59 & 292.83 & F4V & 1.366 & K4V & M0.5V & 0.57 & 11.83 & 351.4 & 0.367 & 0.209 \\
TOI 1161 & 7986 & 9.79 & 9.57 & 500.0 & A6V & 1.826 & K0V & K6V & 0.547 & 20.2 & 600.0 & 0.279 & 0.152 \\
TOI 1162 & 8730 & 9.8 & 9.49 & 352.0 & A2V & 2.033 & G8V & K5V & 0.539 & 14.22 & 422.4 & 0.273 & 0.147 \\
TOI 1163 & 9311 & 9.54 & 9.42 & 148.7 & A1V & 2.189 & G8V & K4V & 0.533 & 6.01 & 178.44 & 0.253 & 0.135 \\
TOI 1170 & 7734 & 10.46 & 10.35 & 880.11 & A7V & 1.733 & K1V & K6V & 0.551 & 35.56 & 1056.13 & 0.285 & 0.157 \\
TOI 1171 & 7550 & 10.57 & 10.31 & 482.76 & A8V & 1.674 & K2V & K7V & 0.555 & 19.51 & 579.31 & 0.278 & 0.155 \\
TOI 1175 & 6229 & 10.64 & 10.2 & 216.52 & F7V & 1.202 & K6V & M1V & 0.572 & 8.75 & 259.82 & 0.369 & 0.211 \\
TOI 1178 & 3897 & 11.71 & 10.14 & 36.72 & M0V & 0.604 & M4V & M5V & 0.614 & 1.48 & 44.06 & 0.403 & 0.247 \\
TOI 1181 & 6122 & 10.58 & 10.08 & 302.82 & F8V & 1.166 & K6V & M1.5V & 0.57 & 12.24 & 363.38 & 0.367 & 0.209 \\
TOI 1183 & 5599 & 10.35 & 9.95 & 112.87 & G6V & 0.995 & M0V & M2.5V & 0.515 & 4.56 & 135.44 & 0.364 & 0.187 \\
TOI 1184 & 4534 & 10.99 & 9.95 & 58.59 & K4V & 0.722 & M2.5V & M4V & 0.566 & 2.37 & 70.31 & 0.415 & 0.235 \\
TOI 1189 & 5286 & 10.42 & 9.87 & 248.6 & K0V & 0.873 & M1V & M3V & 0.528 & 10.05 & 298.32 & 0.368 & 0.194 \\
TOI 1191 & 6800 & 10.22 & 9.78 & 355.81 & F2V & 1.439 & K3V & M0V & 0.563 & 14.38 & 426.97 & 0.273 & 0.154 \\
TOI 1192 & 6479 & 10.94 & 10.5 & 283.03 & F5V & 1.274 & K5V & M0.5V & 0.57 & 11.44 & 339.64 & 0.367 & 0.21 \\
TOI 1195 & 5246 & 11.28 & 10.47 & 500.0 & K0V & 0.86 & M1V & M3V & 0.528 & 20.2 & 600.0 & 0.358 & 0.189 \\
TOI 1196 & 6689 & 10.92 & 10.43 & 424.44 & F3V & 1.417 & K4V & M0V & 0.564 & 17.15 & 509.33 & 0.276 & 0.156 \\
TOI 1197 & 7649 & 11.08 & 10.42 & 405.63 & A8V & 1.7 & K2V & K7V & 0.555 & 16.39 & 486.76 & 0.275 & 0.153 \\
TOI 1201 & 3506 & 12.26 & 10.95 & 37.89 & M2.5V & 0.3 & M5V & M6V & 0.727 & 1.53 & 45.46 & 0.592 & 0.431 \\
TOI 1235 & 3912 & 11.5 & 9.92 & 39.63 & K9V & 0.607 & M3.5V & M5V & 0.604 & 1.6 & 47.56 & 0.406 & 0.245 \\
TOI 1237 & 6212 & 10.77 & 10.28 & 243.25 & F7V & 1.198 & K6V & M1V & 0.572 & 9.83 & 291.9 & 0.369 & 0.211 \\
TOI 1241 & 5826 & 11.71 & 11.16 & 546.52 & G1V & 1.049 & K8V & M2V & 0.52 & 22.08 & 655.83 & 0.356 & 0.185 \\
TOI 1251 & 5273 & 11.51 & 10.65 & 185.95 & K0V & 0.902 & M1V & M3V & 0.528 & 7.51 & 223.14 & 0.369 & 0.195 \\
TOI 1263 & 5098 & 9.36 & 8.53 & 46.55 & K2V & 0.839 & M1.5V & M3.5V & 0.539 & 1.88 & 55.86 & 0.34 & 0.183 \\
TOI 1264 & 5040 & 11.47 & 10.63 & 141.76 & K2V & 0.893 & M1.5V & M3.5V & 0.539 & 5.73 & 170.11 & 0.367 & 0.198 \\
TOI 1265 & 6532 & 10.48 & 10.03 & 341.08 & F5V & 1.34 & K5V & M0.5V & 0.57 & 13.78 & 409.29 & 0.365 & 0.208 \\
TOI 1267 & 6378 & 12.0 & 11.64 & 980.0 & F6V & 1.267 & K5V & M1V & 0.569 & 39.6 & 1176.0 & 0.337 & 0.192 \\
TOI 1287 & 5891 & 9.16 & 8.59 & 92.72 & G1V & 1.108 & K8V & M2V & 0.52 & 3.75 & 111.26 & 0.36 & 0.187 \\
TOI 1288 & 6180 & 10.44 & 9.93 & 114.87 & F8V & 0.999 & K6V & M1.5V & 0.57 & 4.64 & 137.84 & 0.364 & 0.208 \\
TOI 1290 & 5875 & 10.08 & 9.51 & 144.17 & G1V & 1.153 & K8V & M2V & 0.52 & 5.83 & 173.0 & 0.367 & 0.191 \\
TOI 1301 & 4781 & 11.34 & 10.52 & 90.87 & K3V & 0.772 & M2V & M3.5V & 0.548 & 3.67 & 109.04 & 0.419 & 0.23 \\
TOI 1305 & 6400 & 10.79 & 9.98 & 664.0 & F6V & 1.277 & K5V & M1V & 0.569 & 26.83 & 796.8 & 0.35 & 0.199 \\
TOI 1306 & 5273 & 10.69 & 10.02 & 364.48 & K0V & 0.867 & M1V & M3V & 0.528 & 14.73 & 437.38 & 0.364 & 0.192 \\
TOI 1307 & 5009 & 11.61 & 10.8 & 765.3 & K2V & 0.813 & M1.5V & M3.5V & 0.539 & 30.92 & 918.36 & 0.346 & 0.186 \\
TOI 1311 & 8153 & 10.82 & 10.58 & 556.81 & A5V & 1.849 & K0V & K6V & 0.543 & 22.5 & 668.17 & 0.28 & 0.152 \\
TOI 1314 & 5155 & 10.75 & 9.9 & 285.36 & K1V & 0.848 & M1V & M3V & 0.53 & 11.53 & 342.43 & 0.367 & 0.195 \\
TOI 1315 & 8321 & 9.35 & 9.15 & 453.74 & A4V & 1.918 & G9V & K5V & 0.545 & 18.33 & 544.49 & 0.277 & 0.151 \\
TOI 1316 & 6555 & 10.73 & 10.33 & 434.96 & F5V & 1.35 & K5V & M0.5V & 0.57 & 17.58 & 521.95 & 0.361 & 0.206 \\
TOI 1317 & 8542 & 10.7 & 10.29 & 697.12 & A3V & 1.998 & G8V & K5V & 0.541 & 28.17 & 836.54 & 0.283 & 0.153 \\
TOI 1320 & 6500 & 10.73 & 10.33 & 340.29 & F5V & 1.325 & K5V & M0.5V & 0.57 & 13.75 & 408.35 & 0.365 & 0.208 \\
TOI 1321 & 9067 & 10.4 & 10.08 & 1139.32 & A1V & 2.106 & G8V & K4V & 0.533 & 46.04 & 1367.18 & 0.287 & 0.153 \\
TOI 1323 & 9003 & 8.33 & 8.27 & 301.42 & A2V & 2.088 & G8V & K5V & 0.539 & 12.18 & 361.7 & 0.27 & 0.146 \\
TOI 1324 & 6854 & 10.55 & 10.19 & 339.74 & F2V & 1.445 & K3V & M0V & 0.563 & 13.73 & 407.69 & 0.272 & 0.153 \\
TOI 1327 & 8715 & 8.52 & 8.38 & 457.79 & A2V & 2.031 & G8V & K5V & 0.539 & 18.5 & 549.35 & 0.277 & 0.15 \\
TOI 1328 & 6585 & 10.87 & 10.36 & 397.91 & F4V & 1.363 & K4V & M0.5V & 0.57 & 16.08 & 477.49 & 0.363 & 0.207 \\
TOI 1329 & 7537 & 10.55 & 10.28 & 398.63 & A8V & 1.668 & K2V & K7V & 0.555 & 16.11 & 478.36 & 0.275 & 0.153 \\
TOI 1334 & 11348 & 9.66 & 9.55 & 1276.0 & B9V & 3.209 & G1V & K3V & 0.519 & 51.56 & 1531.2 & 0.287 & 0.149 \\
TOI 1339 & 5461 & 8.97 & 8.29 & 53.61 & G8V & 0.939 & M0.5V & M3V & 0.515 & 2.17 & 64.33 & 0.345 & 0.178 \\
TOI 1342 & 5869 & 10.97 & 10.42 & \nodata & G1V & 1.067 & \nodata & \nodata & \nodata & \nodata & \nodata & \nodata & \nodata \\
TOI 1353 & 9706 & 10.25 & 10.15 & 464.21 & A0V & 2.33 & G5V & K4V & 0.531 & 18.76 & 557.05 & 0.278 & 0.147 \\
TOI 1354 & 9224 & 8.83 & 8.79 & 245.78 & A1V & 2.017 & G8V & K4V & 0.533 & 9.93 & 294.94 & 0.266 & 0.142 \\
TOI 1355 & 9218 & 8.72 & 8.65 & 248.15 & A1V & 2.264 & G8V & K4V & 0.533 & 10.03 & 297.78 & 0.266 & 0.142 \\
TOI 1356 & 12308 & 9.04 & 8.95 & 4772.59 & B8V & 3.366 & F6V & K0V & 0.524 & 192.84 & 5727.11 & 0.28 & 0.146 \\
TOI 1357 & 7387 & 10.82 & 10.22 & 254.4 & A9V & 1.548 & K2V & K7V & 0.557 & 10.28 & 305.28 & 0.266 & 0.148 \\
TOI 1358 & 10445 & 8.65 & 8.64 & 356.68 & B9.5V & 2.627 & G2V & K3V & 0.523 & 14.41 & 428.02 & 0.273 & 0.143 \\
TOI 1360 & 8999 & 10.14 & 10.01 & 533.62 & A2V & 2.087 & G8V & K5V & 0.539 & 21.56 & 640.34 & 0.28 & 0.151 \\
TOI 1362 & 8129 & 10.38 & 9.77 & 975.53 & A5V & 1.862 & K0V & K6V & 0.543 & 39.42 & 1170.64 & 0.286 & 0.155 \\
TOI 1364 & 9598 & 9.24 & 9.21 & 595.29 & A0V & 2.278 & G5V & K4V & 0.531 & 24.05 & 714.35 & 0.281 & 0.149 \\
TOI 1366 & 5717 & 9.11 & 8.48 & 118.59 & G3V & 0.999 & K9V & M2.5V & 0.512 & 4.79 & 142.31 & 0.365 & 0.187 \\
TOI 1367 & 6550 & 9.71 & 9.24 & 195.86 & F5V & 1.348 & K5V & M0.5V & 0.57 & 7.91 & 235.03 & 0.369 & 0.21 \\
TOI 1368 & 5717 & 9.78 & 9.14 & 158.18 & G3V & 0.999 & K9V & M2.5V & 0.512 & 6.39 & 189.82 & 0.368 & 0.188 \\
TOI 1369 & 9139 & 10.59 & 10.42 & 516.3 & A1V & 2.224 & G8V & K4V & 0.533 & 20.86 & 619.56 & 0.279 & 0.149 \\
TOI 1370 & 8728 & 9.39 & 9.25 & 275.49 & A2V & 2.061 & G8V & K5V & 0.539 & 11.13 & 330.59 & 0.268 & 0.145 \\
TOI 1376 & 6026 & 10.78 & 10.26 & 240.01 & F9V & 1.133 & K6V & M1.5V & 0.557 & 9.7 & 288.01 & 0.369 & 0.205 \\
TOI 1378 & 8224 & 9.82 & 9.62 & 547.4 & A4V & 1.886 & G9V & K5V & 0.545 & 22.12 & 656.88 & 0.28 & 0.153 \\
TOI 1381 & 9976 & 10.55 & 10.38 & 737.11 & A0V & 2.333 & G5V & K4V & 0.531 & 29.78 & 884.53 & 0.283 & 0.15 \\
TOI 1384 & 5773 & 10.68 & 10.0 & 235.22 & G2V & 1.022 & K9V & M2.5V & 0.514 & 9.5 & 282.26 & 0.369 & 0.189 \\
TOI 1385 & 9400 & 8.54 & 8.44 & 324.35 & A1V & 2.219 & G8V & K4V & 0.533 & 13.11 & 389.22 & 0.271 & 0.145 \\
TOI 1386 & 5769 & 10.61 & 10.06 & 146.86 & G2V & 1.019 & K9V & M2.5V & 0.514 & 5.93 & 176.23 & 0.367 & 0.189 \\
TOI 1387 & 7976 & 9.17 & 8.96 & 239.69 & A6V & 1.776 & K0V & K6V & 0.547 & 9.69 & 287.63 & 0.265 & 0.145 \\
TOI 1391 & 5256 & 11.14 & 10.46 & 115.75 & K0V & 0.854 & M1V & M3V & 0.528 & 4.68 & 138.9 & 0.365 & 0.192 \\
TOI 1393 & 7300 & 9.79 & 9.52 & 278.8 & F0V & 1.629 & K2V & K8V & 0.56 & 11.27 & 334.56 & 0.268 & 0.15 \\
TOI 1394 & 6294 & 9.93 & 9.44 & 195.02 & F6V & 1.232 & K5V & M1V & 0.569 & 7.88 & 234.02 & 0.369 & 0.21 \\
TOI 1397 & 6357 & 10.64 & 10.18 & 194.99 & F6V & 1.184 & K5V & M1V & 0.569 & 7.88 & 233.99 & 0.369 & 0.21 \\
TOI 1398 & 9623 & 8.68 & 8.63 & 648.4 & A0V & 2.284 & G5V & K4V & 0.531 & 26.2 & 778.08 & 0.282 & 0.15 \\
TOI 1399 & 6483 & 10.59 & 10.21 & 252.4 & F5V & 1.293 & K5V & M0.5V & 0.57 & 10.2 & 302.88 & 0.368 & 0.21 \\
TOI 1400 & 6004 & 11.55 & 10.94 & 370.79 & F9V & 1.122 & K6V & M1.5V & 0.557 & 14.98 & 444.95 & 0.364 & 0.203 \\
TOI 1401 & 6403 & 11.53 & 10.98 & 383.17 & F6V & 1.302 & K5V & M1V & 0.569 & 15.48 & 459.8 & 0.364 & 0.207 \\
TOI 1402 & 7111 & 10.8 & 10.59 & 576.75 & F1V & 1.536 & K3V & K9V & 0.561 & 23.3 & 692.1 & 0.281 & 0.157 \\
TOI 1405 & 5195 & 9.4 & 8.42 & \nodata & K1V & 0.853 & \nodata & \nodata & \nodata & \nodata & \nodata & \nodata & \nodata \\
TOI 1407 & 6129 & 8.24 & 7.73 & 80.9 & F8V & 1.168 & K6V & M1.5V & 0.57 & 3.27 & 97.08 & 0.357 & 0.204 \\
TOI 1554 & 5497 & 12.02 & 11.06 & 194.03 & G8V & 0.951 & M0.5V & M3V & 0.515 & 7.84 & 232.84 & 0.369 & 0.19 \\
TOI 1905 & 4251 & 11.59 & 10.42 & 64.74 & K6V & 0.697 & M3V & M4.5V & 0.571 & 2.62 & 77.69 & 0.417 & 0.238 \\
TIC125192758 & 5384 & 15.11 & 14.15 & 1100.46 & G9V & 0.914 & M0.5V & M3V & 0.522 & 44.47 & 1320.55 & 0.332 & 0.173 \\
\enddata
\tablenotetext{a}{Spectral type from the Modern Mean Dwarf Stellar Color and Effective Temperature Sequence based on T$_{\mathrm{eff}}$.}
\tablenotetext{b}{Mass interpolated from the Modern Mean Dwarf Stellar Color and Effective Temperature Sequence using R$_{*}$, if available, and T$_{\mathrm{eff}}$ .}
\tablenotetext{c}{${T_\mathrm{eff}}$ not given in ExoFOP, but was estimated using the available magnitudes.}
\end{deluxetable*}



\bibliographystyle{aasjournal.bst}
\bibliography{TESS}



\appendix
\begin{figure}[h!] 
\centering
  \includegraphics[scale=0.85,keepaspectratio=true]{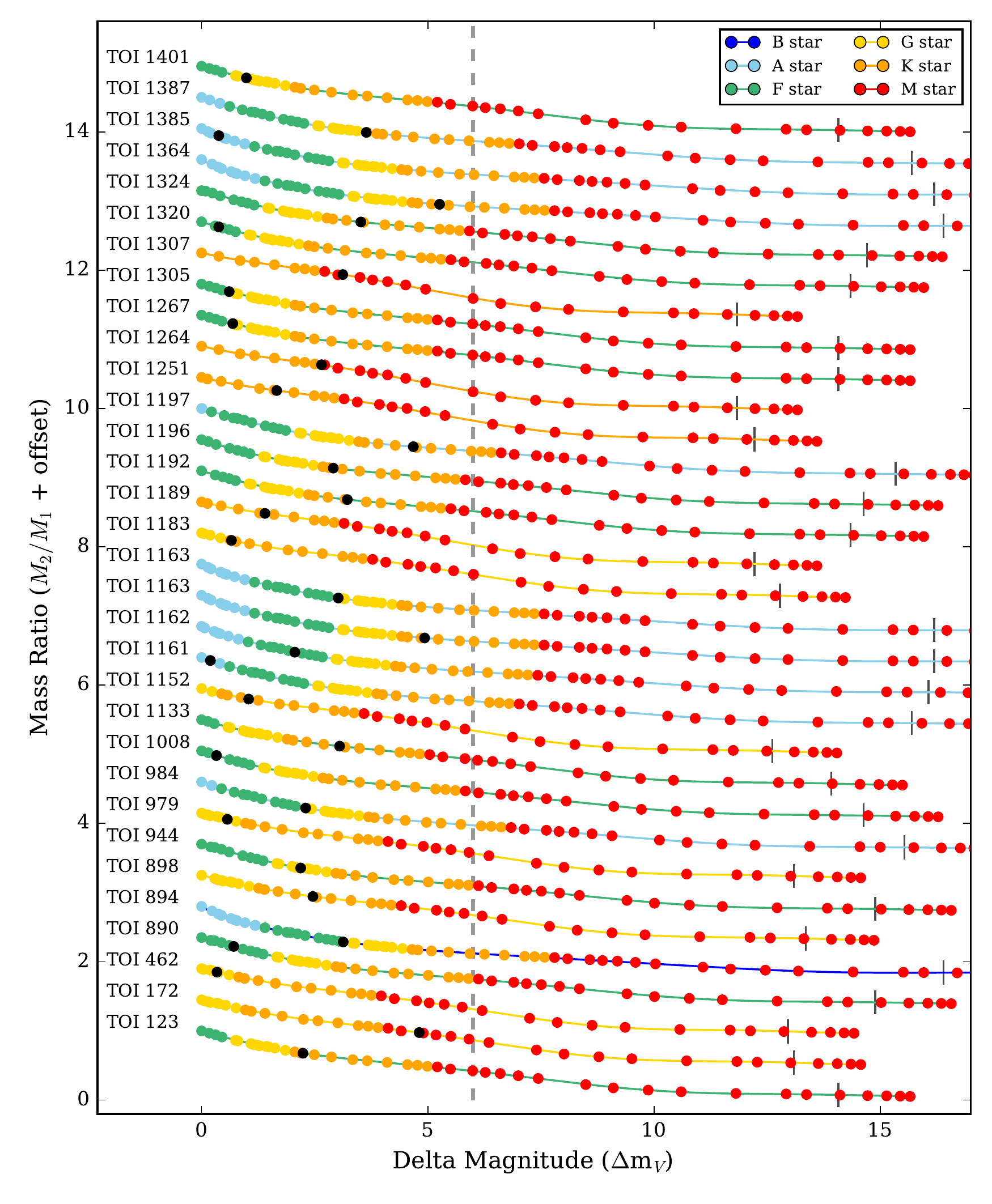}
  \caption{Expected mass ratio (offset for clarity) vs.~$V$-band delta magnitude for TOIs with companions detected at 562nm. The colors of the lines and dots correspond to the spectral type of the star/possible companion. Contrast limits for speckle imaging with NESSI are shown by the dashed line ($\Delta m \lesssim 6$). The small vertical lines along each line indicate where the mass ratio equals 0.1. The black dots show the measured delta magnitude of the observed companions.}
  \centering
\end{figure}

\begin{figure}[h!] 
\centering
  \includegraphics[scale=0.85,keepaspectratio=true]{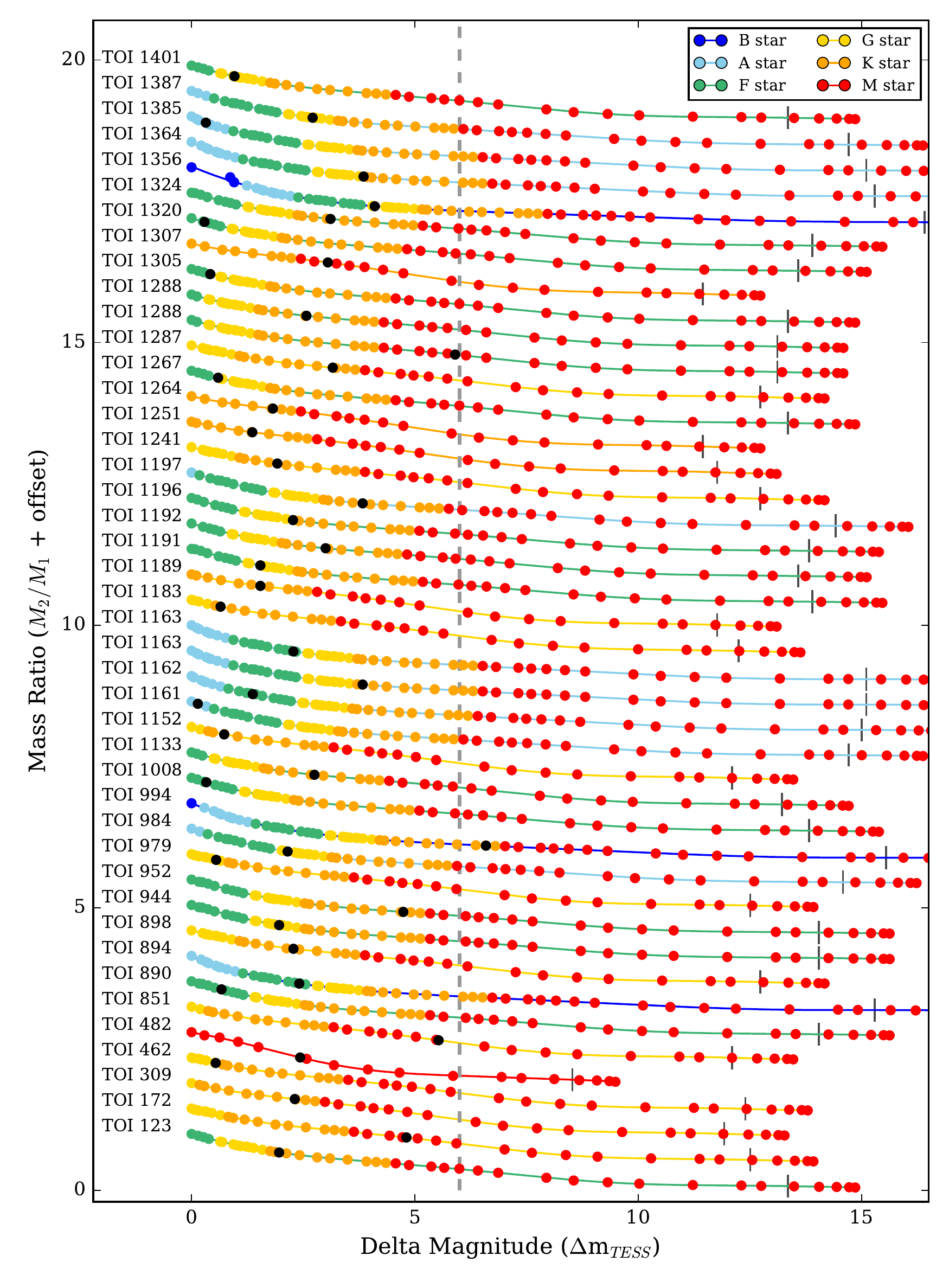}
  \caption{Expected mass ratio (offset for clarity) vs.~TESS-band delta magnitude for TOIs with companions detected at 832nm. The colors of the lines and dots correspond to the spectral type of the star/possible companion. Contrast limits for speckle imaging with NESSI are shown by the dashed line ($\Delta m \lesssim 6$). The small vertical lines along each line indicate where the mass ratio equals 0.1. The black dots show the measured delta magnitude of the observed companions.}
  \centering
\end{figure}

\begin{figure}[h!] 
\centering
  \includegraphics[scale=0.38,keepaspectratio=true]{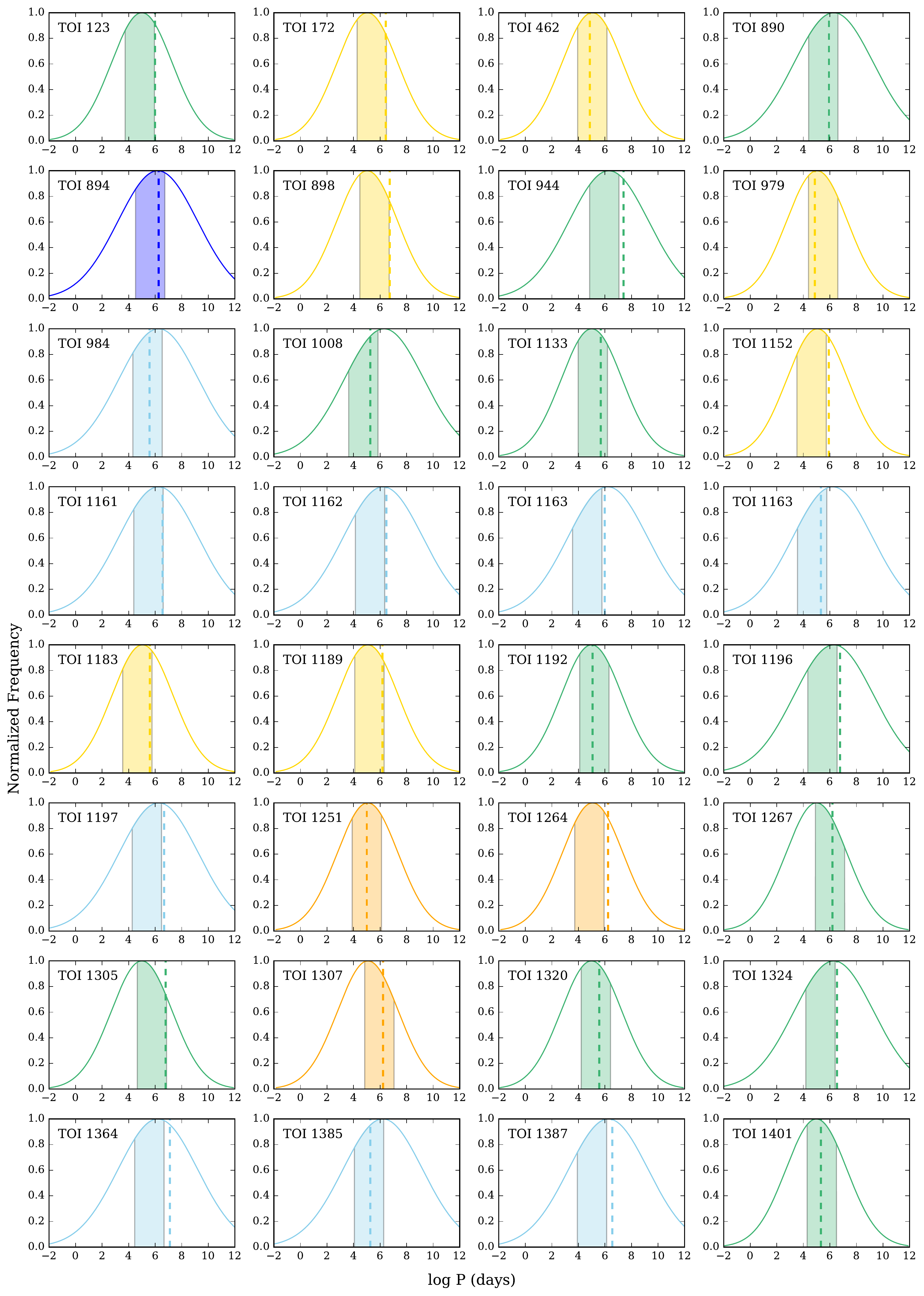}
  \caption{Expected binary period distributions for all TOIs with companions detected in 562nm based on the distribution presented in \citet{2010ApJS..190....1R}. The shaded regions (color coded by spectral type of the primary as in Figure \ref{fig:compfrac}) show the orbital periods corresponding to projected separations at which speckle imaging at WIYN can detect companions ($0.04-1.2"$). The dashed lines correspond to the separation of observed companions converted to $\log P$ space using the distance and mass of the primary star.}
  \centering
\end{figure}

\begin{figure}[h!] 
\centering
  \includegraphics[scale=0.38,keepaspectratio=true]{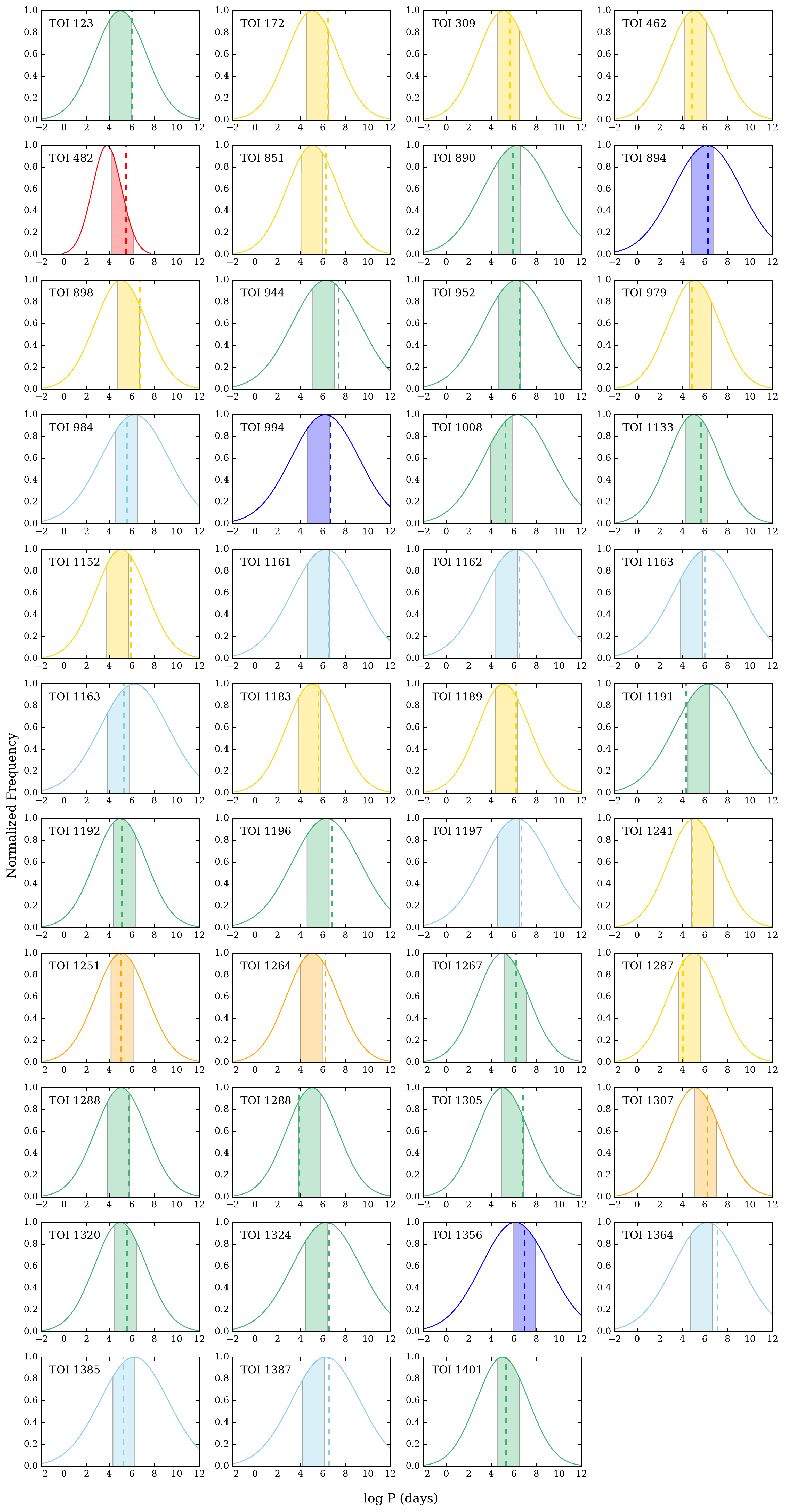}
  \caption{Expected binary period distributions for all TOIs with companions detected in 832nm based on the distribution presented in \citet{2010ApJS..190....1R}. The shaded regions (color coded by spectral type of the primary as in Figure \ref{fig:compfrac}) show the orbital periods corresponding to projected separations at which speckle imaging at WIYN ($0.06-1.2"$) can detect companions. The dashed lines correspond to the separation of observed companions converted to $\log P$ space using the distance and mass of the primary star.}
  \centering
\end{figure}



\end{document}